\documentclass[aps,pra,twocolumn,superscriptaddress]{revtex4}%
\usepackage{graphics}
\usepackage{amsmath}
\usepackage{graphicx}
\usepackage{amsfonts}
\usepackage{amssymb}
\usepackage{float}
\usepackage{longtable}
\usepackage{epsfig}
\usepackage{latexsym}
\usepackage{theorem}
\usepackage{bbm}
\usepackage{bm}
\usepackage{braket}
\usepackage{psfrag}%
\newcommand{\im}{\mathrm{i}}
\newtheorem{theorem}{Theorem}
\newtheorem{lemma}[theorem]{Lemma}
\newtheorem{proposition}[theorem]{Proposition}
\newtheorem{corollary}[theorem]{Corollary}
\newtheorem{definition}[theorem]{Definition}

\begin{document}
\title{Simulation of non-Pauli Channels}
\author{Thomas P. W. Cope}
\affiliation{Computer Science \& York Centre for Quantum
Technologies, University of York, York YO10 5GH, UK}
\author{Leon Hetzel}
\affiliation{Fachbereich 1 Physik $\&$ Elektrotechnik,
Universit\"{a}t Bremen, 28359 Bremen, Germany}
\author{Leonardo Banchi}
\affiliation{Department of Physics and Astronomy, University
College London, Gower Street, London WC1E 6BT, United Kingdom}
\author{Stefano Pirandola}
\affiliation{Computer Science \& York Centre for Quantum Technologies, University of York,
York YO10 5GH, UK}

\begin{abstract}
We consider the simulation of a quantum channel by two parties who
share a resource state and may apply local operations assisted by
classical communication (LOCC). One specific type of such LOCC is
standard teleportation, which is however limited to the simulation
of Pauli channels. Here we show how we can easily enlarge this
class by means of a minimal perturbation of the teleportation
protocol, where we introduce noise in the classical communication
channel between the remote parties. By adopting this noisy
protocol, we provide a necessary condition for simulating a
non-Pauli channel. In particular, we characterize the set of
channels that are generated assuming the Choi matrix of an
amplitude damping channel as a resource state. Within this set, we
identify a class of Pauli-damping channels for which we bound the
two-way quantum and private capacities.
\end{abstract}
\maketitle

\section{Introduction}
Simulation of quantum channels is a central tool in quantum
information theory~\cite{NielsenChuang,Preskill,RMP,SamRMPm}. One
of the first seminal ideas was introduced in Ref.~\cite{B2}, where
the channel simulation was based on the standard teleportation
protocol~\cite{tele,teleREVIEW}, but where the shared
maximally-entangled state was replaced by an arbitrary two-qubit
resource state. Later on, Ref.~\cite{BoBo} showed that this method
allows one to simulate any Pauli channel, i.e., any quantum
channel whose action on an input state can be expressed by a Kraus
decomposition in terms of Pauli operators~\cite{NielsenChuang}. In
Ref.~\cite{B2}, the teleportation simulation was used to transform
protocols of quantum communication through a (Pauli) channel into
protocols of entanglement distillation over the resource states.
The same technique was then exploited in Ref.~\cite{HoroTEL} to
show the reproducibility between (isotropic) states and (Pauli)
channels.

In 2001, Ref.~\cite{WernerTELE} described generalized
teleportation protocols in the context of discrete variable (DV)
systems, allowing for more general quantum measurements beyond
Bell detection. Following these steps, Ref.~\cite{Leung} moved the
first steps in investigating teleportation-covariance for DV
channels, which is that property of a quantum channel to commute
with the random unitaries of teleportation. This property has been
generalized by Ref.~\cite{Stretching} to quantum channels at any
dimension, including continuous variable (CV) channels. Thanks to
teleportation covariance, a quantum channel can be simulated by
teleporting over its Choi matrix. This result was re-stated in a
different form by a follow-up work~\cite{WildeFollowup}.

One crucial step introduced by Ref.~\cite{Stretching} has been the
removal of any restriction on the dimension of the quantum systems
involved in the simulation process. For this reason, one can
simulate DV channels, CV channels and even hybrid channels between
DVs and CVs. More generally, Ref.~\cite{Stretching} was not
limited to teleportation LOCCs (i.e., Bell detection and unitary
corrections), but considered completely general LOCCs which may
also be asymptotic, i.e., defined as suitable sequences. This more
general LOCC simulation allowed them to simulate \textit{any}
quantum channel. In particular, it allowed them to simulate, for
the first time in the literature, the amplitude damping channel
(which is a DV channel) by using the Choi matrix of a bosonic
lossy channel (which is a CV channel) and an LOCC based on hybrid
CV-DV teleportation maps~\cite{Nota1}.


One of the most powerful applications of channel simulation is
teleportation stretching~\cite{Stretching}. In this method, the
LOCC simulation of a quantum channel (with some resource state
$\sigma$) is used to completely simplify the structure of adaptive
protocols of quantum and private communication, which are based on
the use of adaptive LOCCs, i.e., local operations assisted by
unlimited and two-way classical communications (CCs). Any such
protocol can be re-organized in such a way to become a much
simpler block protocol, where the output state, after $n$ uses of
the channel, is expressed in terms of a tensor-product of the
resource states $\sigma^{\otimes n}$ up to a global LOCC. Contrary
to previous approaches~\cite{B2,Niset,Wolfnotes,AlexAPP}, the
method devised in Ref.~\cite{Stretching} does not reduce quantum
communication (over specific channels) into entanglement
distillation, but reduce \textit{any} adaptive protocol (over
\textit{any} channel at \textit{any} dimension) into an equivalent
block form, where the original task is perfectly preserved (e.g.,
so that adaptive key generation is transformed into block key
generation). For this reason, the technique has been also extended
beyond point-to-point quantum
communication~\cite{networkPIRS,Multipointa}, and also to simplify
adaptive protocols of quantum metrology and quantum channel
discrimination~\cite{Metro,Nota2}.


By using teleportation stretching and extending the notion of
relative entropy of entanglement
(REE)~\cite{RMPrelent,VedFORMm,Pleniom} from states to channels,
Ref.~\cite{Stretching} derived a simple single-letter bound for
the two-way quantum and private capacities of an arbitrary quantum
channel. Such bound is shown to be achievable in many important
cases, so that Ref.~\cite{Stretching} established these capacities
for dephasing channels, erasure channels (see also
Refs.~\cite{ErasureChannel,GEWa}), quantum-limited amplifiers, and
bosonic lossy channels. The two-way capacity of the lossy channel,
also known as Pirandola-Laurenza-Ottaviani-Banchi (PLOB) bound,
completes an investigation started back in
2009~\cite{RevCohINFO,ReverseCAP}, and finally sets the ultimate
achievable limit for optical quantum communications in the absence
of quantum repeaters. This benchmark for quantum repeaters has
been already exploited in
literature~\cite{bench1,bench2,bench3,bench4,bench5}. Building on
most of the methods discovered by Ref.~\cite{Stretching} (i.e.,
channel's REE and teleportation stretching), the follow-up
work~\cite{WildeFollowup} later discussed the strong converse
property of the various bounds and two-way capacities established
in Ref.~\cite{Stretching}. See Ref.~\cite{Nota3} for
clarifications on literature.

In this context, the present work brings several new insights. It
considers a minimum perturbation of the standard teleportation
protocol, where the noiseless classical communication channel
between the parties (Alice and Bob) is replaced by a noisy
classical channel, where the Bell outcomes $k$ are stochastically
mapped into a variable $l$ on the same alphabet, according to some
conditional probability distribution $p_{l|k}$. We show that this
already allows us to enlarge the class of simulable channels well
beyond that of Pauli channels. This is non-trivial because this is
achieved without changing the dimensions of Alice's and Bob's
local Hilbert spaces $\mathcal{H}_A$ and $\mathcal{H}_B$
associated with the resource state $\sigma=\sigma_{AB}$. In fact,
changing such dimensions is another way to generate non-Pauli
channels, an example being the erasure channel which can be
generated using a $2\times 3$ dimensional resource state (i.e., a
qubit entangled with a qutrit).

Adopting the vectorial Bloch sphere representation for
qubits~\cite{NielsenChuang}, we provide simple conditions to be
satisfied in order to simulate non-Pauli channels. A profitable
way to generate such kinds of channels is to start from the Choi
matrix of an amplitude damping channel as resource state for the
noisy teleportation protocol. In this way, we can generate
non-Pauli channels which are significantly far from the Pauli
class, as quantified by the trace norm and the diamond norm.
In particular, we identify a class of simulable channels that we
call ``Pauli-damping channels'' because they can be decomposed
into a Pauli and an amplitude damping part. For channels in this
class we compute lower and upper bounds for the two-way quantum
and private capacities, by adopting the methodology developed by
Ref.~\cite{Stretching}.

The paper is structured as follows. We start with discussing
preliminary notions in Sec.~\ref{SEC_preli}, including the basics
of quantum teleportation, channel simulation and teleportation
stretching and its application to derive upper bounds for the
two-way capacities. Then, in Sec.~\ref{Expanding}, we show how to
simulate non-Pauli channels via our noisy teleportation protocol.
This is further developed in Sec.~\ref{dampSEC}, where we consider
the channels simulated starting from the Choi matrix of the
amplitude damping channel and we also define the Pauli-damping
channels. The properties of these channels are studied in
Sec.~\ref{SECprop}. Finally, Sec.~\ref{SECconclu} is for
conclusions.

\section{Preliminaries}\label{SEC_preli}
\subsection{Quantum teleportation}\label{QuanTel}
Teleportation~\cite{tele,TeleCon,teleREVIEW,teleMORE1,teleMORE2,teleMORE3}
is one of the strangest and most intriguing results to come out of
quantum information. We shall outline the standard approach here,
so that the generalizations in the following sections are more
apparent. The basic version of the protocol is as follows. Alice
($A$) and Bob ($B$) share a maximally entangled state, e.g., a
Bell state of the form
\begin{equation}\label{maxENT}
\ket{\Phi}=\frac{1}{\sqrt{d}}\left(\sum_{i=0}^{d-1}\ket{i}_A\ket{i}_B\right)
\end{equation}
for DV systems, and the asymptotic EPR state~\cite{SamRMPm}
\begin{equation}
\lim_{r\rightarrow\infty}\sqrt{1-\mathrm{tanh}^2(r)}\sum_{n=0}^\infty[-\mathrm{tanh(r)}]^n\ket{n}_A\ket{n}_B
\end{equation}
for CV systems (where $\ket{n}$ is the number state), which
produces correlations $\hat{q}_A=\hat{q}_B$ for the
position-quadrature, and $\hat{p}_A=-\hat{p}_B$ for the
momentum-quadrature~\cite{TeleCon,teleMORE1}. For the qubit case
$d=2$ which we shall be focusing on later, the state of
Eq.~(\ref{maxENT}) is
$\frac{1}{\sqrt{2}}\left(\ket{00}+\ket{11}\right)$.

Alice also has an arbitrary state $\rho_C$ to be teleported to
Bob. To begin the process, Alice performs a Bell measurement on
her two systems, $AC$. In DVs this is done by using the $d$
dimensional Bell basis, consisting of the $d^2$ maximally
entangled states $\ket{\Phi_{\alpha,\beta}}$, with
$\alpha,\beta\in\left\{1\ldots\;d\right\}$. In operator notation
we describe the measurement by $\left\{M_{\alpha,\beta}\right\}$,
with
$M_{\alpha,\beta}=\ket{\Phi_{\alpha,\beta}}\bra{\Phi_{\alpha,\beta}}$
and
\begin{equation}
 \ket{\Phi_{\alpha,\beta}}
 \equiv(\mathbb{I}_d\otimes\sigma_x^\alpha\sigma_z^\beta)\ket{\Phi},
 \end{equation}
 where
 \begin{align}
 \sigma_x\ket{k}=\ket{k + 1\text{ mod }d},&&\sigma_z\ket{k}=\omega^k\ket{k},\;\;\;\omega=e^{\frac{\im 2\pi}{d}}.
 \end{align}
 The set $\{\sigma_x^\alpha\sigma_z^\beta\}$ is known as the $d$-dimensional Weyl-Heisenberg group.
 In the qubit case, we use the usual set of Pauli operators~\cite{NielsenChuang}
\begin{align}
\mathrm{I}&=\left(\begin{array}{cc}
\phantom{-}1 & \phantom{-}0\\
\phantom{-}0 & \phantom{-}1
\end{array}\right)& \sigma_x&=\left(\begin{array}{cc}
\phantom{-}0 & \phantom{-}1\\
\phantom{-}1 & \phantom{-}0
\end{array}\right)\\
\im\sigma_y\equiv\sigma_x\sigma_z&=\left(\begin{array}{cc}
\phantom{-}0 & \phantom{-}1\\
-1 & \phantom{-}0
\end{array}\right)& \sigma_z&=\left(\begin{array}{cc}
\phantom{-}1 & \phantom{-}0\\
\phantom{-}0 & -1
\end{array}\right).
\end{align}
In CVs, the measurement operator can be thought of as
 \begin{equation}
 M_{k}=(\mathbb{I}\otimes\hat{D}(k))\ket{\Phi}\bra{\Phi}(\mathbb{I}\otimes\hat{D}(k))^\dagger
 \end{equation}
 where $D(k)=\textrm{exp}{(k\hat{a}^\dagger-k^*\hat{a})}$ is the displacement operator with complex amplitude $k$ and $\hat{a}$ being the annihilation operator.

The effect of the Bell measurement, in which all outcomes occur
with equal probability, is to transform Bob's half of the
maximally entangled state into the teleported state up to a random
unitary. In DVs, the state of Bob (system $B$) takes the form
$\rho_{B|(\alpha,\beta)}=\sigma_x^\alpha\sigma_z^\beta\rho_C(\sigma_x^\alpha\sigma_z^\beta)^\dagger$,
for a given Bell outcome $(\alpha,\beta)$, while for CVs this
state is $\rho_{B|k}=\hat{D}(k)\rho_C\hat{D}(k)^\dagger$, given
the Bell outcome $k$. Since Alice communicates the Bell outcome to
Bob, he can undo the random unitary and recover Alice's input
state $\rho_C$. Note that the Alice's CC to Bob is necessary to
reproduce the state, otherwise the two remote users could
communicate faster than the speed of light. In the following, we
focus on DV systems and we discuss how the teleportation protocol
can be progressively modified to simulate more and more quantum
channels.

\subsection{Changing the resource for teleportation}\label{ResChange}
From the protocol described in the previous section, a natural
question to ask is ``what is the consequence of changing the
resource state shared by Alice and Bob?" This was first considered
in~\cite{B2}, who looked into the scenario where Alice and Bob
instead share a generic mixed two-qubit state, which we can
express as \cite{Horodecki}
\begin{align}
\tau=\frac{1}{4}\bigg(&\mathrm{I}\otimes\mathrm{I}+\sum_{i=1}^3 a_i\sigma_i\otimes\mathrm{I}\nonumber\\
+&\sum_{j=1}^3\mathrm{I}\otimes b_j\sigma_j +
\sum_{i,j=1}^{3}t_{ij}\sigma_i\otimes
\sigma_j\bigg)\label{generaltwo}.
\end{align}
in terms of Pauli operators
$\{\sigma_i\}_{i=0}^3=\{I,\sigma_x,\sigma_y,\sigma_z\}$, the
vectors  $\mathbf{a}=\{a_{i}\}$, $\mathbf{b}=\{b_{i}\}$, and the
matrix $[T]_{ij}=t_{ij}$.

\begin{theorem}[\cite{BoBo}]\label{Bobothe}
The effect of teleportation over an arbitrary two-qubit state
$\tau$ as in Eq.~(\ref{generaltwo}) is the Pauli channel
\begin{equation}\label{Pauliform}
\mathcal{E}_P:\rho\rightarrow\sum_{i=0}^3 p_i\sigma_i\rho\sigma_i,
\end{equation}
where $p_i=\mathrm{Tr}\left(E_i\tau\right)$ and $E_i$ are the
projectors on the Bell states, i.e.,
\begin{align}
E_0&=\ket{\Phi^+}\bra{\Phi^+},~~\ket{\Phi^+}:=\frac{1}{\sqrt{2}}\left(\ket{00}+\ket{11}\right),\\
E_1&=\ket{\Psi^+}\bra{\Psi^+},~~\ket{\Psi^+}:=\frac{1}{\sqrt{2}}\left(\ket{01}+\ket{10}\right),\\
E_2&=\ket{\Psi^-}\bra{\Psi^-},~~\ket{\Psi^-}:=\frac{1}{\sqrt{2}}\left(\ket{01}-\ket{10}\right),\\
E_3&=\ket{\Phi^-}\bra{\Phi^-},~~\ket{\Phi^-}:=\frac{1}{\sqrt{2}}\left(\ket{00}-\ket{11}\right).
\end{align}
\end{theorem}

Using this theorem, we can view the standard teleportation
protocol of Sec.~\ref{QuanTel} in a new context, as simulating a
trivial Pauli channel (the identity channel from Alice to Bob). We
can re-state the previous theorem by using the \textit{Bloch
sphere representation} of qubit states.

\begin{definition}[\cite{NielsenChuang}]
In the computational basis, an arbitrary qubit state $\rho$ can be
represented by the density matrix\begin{equation} \rho
=\frac{1}{2}\left(\begin{array}{cc}
1+ z & x - \im y \\
x + \im y & 1 - z
\end{array}
\right).
\end{equation}
This is one-to-one with a Bloch vector, $\mathbf{r}=(x,y,z)$, with
Euclidean norm $||\mathbf{r}||\leq 1$ (equality for pure states).
We can thus represent the actions of qubit channels by their
effect on the Bloch vector of the sent state.
\end{definition}

Given a generic resource state of the form (\ref{generaltwo}), we
easily find that the Pauli channel simulated by teleportation over
this state corresponds to the transformation
\begin{equation}\label{TRASF}
\mathcal{E}:\left(x,y,z\right)\rightarrow\left(t_{11}x,-t_{22}y,t_{33}z\right),
\end{equation}
of the Pauli channel as follows
\begin{align}
t_{11}&=\phantom{-}p_0+p_1-p_2-p_3&&=\phantom{-}1-2p_2-2p_3,\\
t_{22}&=-p_0+p_1-p_2+p_3&&= -1+2p_1+2p_3,\\
t_{33}&=\phantom{-}p_0-p_1-p_2+p_3&&=\phantom{-}1-2p_1-2p_2.
\end{align}
It is also easy to verify that
\begin{align}
t_{11}+t_{22}+t_{33}&\leq 1,\\
t_{11}-t_{22}-t_{33}&\leq 1, \\
-t_{11}+t_{22}-t_{33}&\leq 1,\\
-t_{11}-t_{22}+t_{33}&\leq 1,
\end{align}
which means that the vector $(t_{11},t_{22},t_{33})$,
characterizing the Pauli channel, must belong to the tetrahedron
$\mathcal{T}$ defined by the convex combination of the four points
\begin{align}\label{tetra}
\mathbf{e}_0&=(\phantom{-}1,-1,\phantom{-}1),&\mathbf{e}_1&=(\phantom{-}1,\phantom{-}1,-1),\\
\mathbf{e}_2&=(-1,-1,-1),&\mathbf{e}_3&=(-1,\phantom{-}1,\phantom{-}1).\nonumber
\end{align}

According to Eq.~(\ref{TRASF}), there is a simple way to simulate
a Pauli channel with arbitrary probability distribution
$\left\{p_i\right\}$. One may just take the resource state
\begin{equation}\label{PauliChoi}
\rho=\frac{1}{4}\left(\mathrm{I}\otimes\mathrm{I}+\sum_{i=1}^3t_{ii}\sigma_i\otimes\sigma_i\right),
\end{equation}
with $t_{ii}$ being connected to $\left\{p_i\right\}$ by the
formulas above. Note that this resource state is \textit{Bell
diagonal}, i.e., a mixture of the four Bell states.

\subsection{Generalized channel simulation}\label{Simulation}
In general, the simulation of a quantum channel does not
necessarily need to be implemented through quantum teleportation
(even in some generalized form~\cite{WernerTELE}). In fact, we may
consider a completely arbitrary LOCC applied to some resource
state~\cite{NoteSIM}.

\begin{definition}[\cite{Stretching}]
A quantum channel $\mathcal{E}$ is called $\tau $-stretchable if
there exists an LOCC $\mathcal{S}$ and a resource state $\tau$
simulating the channel. More precisely, for any input state
$\rho$, we may write
\begin{equation}\label{ggg}
\mathcal{E}(\rho)=\mathcal{S}(\rho\otimes\tau)~.
\end{equation}
\end{definition}

Note that this is an extremely general idea. The dimension of the
Hilbert spaces involved can be finite, infinite, equal or
non-equal. Because of the generality of the LOCC, it is clear that
any channel is (trivially) simulable by a maximally entangled
state. In fact, it is sufficient to include the channel
$\mathcal{E}$ into Alice's LOs and then perform the standard
teleportation of the output. In fact, the point is to find the
best resource state $\tau$ among all the possible LOCC
simulations. Typically, the best case is when $\tau$ represents
the Choi matrix of channel
\begin{equation}
\chi_\mathcal{E}:=\mathbb{I}\otimes\mathcal{E}\left(\ket{\Phi}\bra{\Phi}\right).
\end{equation}

\begin{definition}[\cite{Stretching}]
A quantum channel $\mathcal{E}$ is called \textquotedblleft
Choi-stretchable" if it can be LOCC-simulated by using its Choi
matrix, i.e., we can write Eq.~(\ref{ggg}) with
$\tau=\chi_{\mathcal{E}}$.
\end{definition}

There is a simple condition that allows us to identify
Choi-stretchable channels, teleportation covariance.

\begin{definition}[\cite{Stretching}]
A quantum channel $\mathcal{E}$ is called
``teleportation covariant'' if, for any teleportation unitary $U$,
there exists some unitary $V$ such that
\begin{equation}
\mathcal{E}\left(  U\rho U^{\dagger}\right)  =V\mathcal{E}\left(
\rho\right)
V^{\dagger}.\label{covdef}%
\end{equation}
\end{definition}

Because of teleportation covariance we can simulate a quantum
channel by means of teleportation over its Choi matrix. In fact, let $\rho_{C}$ be an input state (owned by Alice) of channel $\mathcal{E}$ and consider the teleportation of $\rho_{C}$ using the maximally
entangled state $\ket{\Phi}_{AB}$. When Alice performs her Bell
measurement, if the outcome corresponding to the Bell state $(\mathbb{I}\otimes U)\ket{\Phi}$ is obtained, then the
state $U\rho_{C}U^\dagger$ is teleported to $B$. Applying a
teleportation covariant $\mathcal{E}$ to this state, we obtain
\begin{equation}
\mathcal{E}\left(U\rho_{C}U^\dagger\right)=V\mathcal{E}\left(\rho_{C}\right)V
^\dagger.
\end{equation}
Therefore, if the corrective unitary $V^{-1}$ is applied by Bob \textit{after} the channel for all the possible $U$, then he will obtain the final state $\mathcal{E}\left(\rho_{C}\right)$ irrespective of the Bell detection outcome.
This corresponds to simulation of $\mathcal{E}$ by teleportation. However,
because the Bell measurement on systems $AC$ is locally separated
from the application of $\mathcal{E}$ on system $B$, we can
commute these operations and the result is the simulation of
$\mathcal{E}$ by teleporting over its Choi matrix
$\chi_\mathcal{E}$. This leads to the following.
\begin{lemma}[\cite{Stretching}]
If a quantum channel $\mathcal{E}$ is teleportation covariant,
then it is Choi-stretchable via teleportation. This channel may
also be called a ``teleportation simulable'' channel.
\end{lemma}
All Pauli channels (regardless of dimension) are teleportation
covariant, and are therefore Choi-stretchable.

Note that in the previous lemma, we are stating a sufficient
condition only. We would like to modify the lemma into a
sufficient and necessary condition. Let us define the
Weyl-Heisenberg (WH) teleportation protocol. This is a
teleportation protocol over an arbitrary resource state where the
output corrective unitary is a unitary representation of the
Weyl-Heisenberg group associated with the Bell detection. This
protocol defines the WH-teleportation channels as follows.

\begin{definition}
We say that a quantum channel is a \textquotedblleft
WH-teleportation channel\textquotedblright\ if it can be written
in the form
\begin{equation}
\Gamma_{\tau}(\rho):=\sum_{g\in G}V_{B}^{\dagger}(g)\mathrm{Tr}_{CA}%
[E_{CA}(g)(\rho_{C}\otimes\tau_{AB})]V_{B}(g),\label{WHform}%
\end{equation}
where $\tau_{AB}$ is a preshared resource state between Alice and
Bob, $E_{CA}(g)=U_{A}^{\dagger}(g)\ket{\Phi}\bra{\Phi}U_{A}(g)$ is
a Bell detection operator with $U(g)\in\left\{
\sigma_{x}^{\alpha}\sigma_{y}^{\beta}\right\} $ belonging to the
$d$-dimensional Weyl-Heisenberg group, and $V(g)$ is a
(generally different) representation of the same group.
\end{definition}

Note that conventional teleportation may be written in the form of
Eq.~(\ref{WHform}) by setting $V(g)=U(g)$ and
$\tau_{AB}=\ket{\Phi}\bra{\Phi}$, the maximally entangled state.
In Appendix~\ref{App:proof1}, we then show the following
characterization.

\begin{theorem}\label{WHtheorem}
For DV systems, a channel is teleportation covariant iff it is a
WH-teleportation channel, i.e., Choi-stretchable via a
WH-teleportation protocol.
\end{theorem}

\subsection{Teleportation stretching and weak converse bounds for private
communication\label{Adaptive}}

The most general protocol for key generation (or private
communication) between two remote parties, connected by a quantum
channel $\mathcal{E}$, consists in the use of adaptive LOCCs
interleaved between each transmission through the channel. This
type of private protocol is very difficult to study due to the
presence of feedback that may be exploited to improve the inputs
to the channel in a real-time fashion. As Ref.~\cite{Stretching} has recently shown,
an adaptive protocol for private communication can be transformed
into a much simpler (non-adaptive) protocol by means of
teleportation stretching. This means that each use of channel
$\mathcal{E}$ is replaced by its\ simulation via an LOCC and a
corresponding resource state $\tau$. All the LOCCs, both the
original from the protocol and the new ones introduced by the
simulation, can be collapsed into a single (trace-preserving) LOCC
$\Lambda$. As a result, after $n$ transmissions, the output of the
protocol can be decomposed into the form
\begin{equation}
\rho_{n}=\Lambda(\tau^{\otimes n}).\label{LOCCsim}%
\end{equation}

To understand the huge simplification that this method brings, we
need to combine it with the use of the relative entropy of
entanglement (REE)~\cite{RMPrelent,VedFORMm,Pleniom}. Recall that
the relative entropy between two states $\rho$ and $\sigma$ is
defined as~\cite{RMPrelent}
\begin{equation}
S(\rho||\sigma):=\mathrm{Tr}(\rho\log\rho-\rho\log\sigma),
\end{equation}
and the REE of a state is given by the following minimization over
all separable states (SEP)~\cite{VedFORMm,Pleniom}
\begin{equation}
E_{R}\left(  \rho\right)
:=\min_{\sigma\in\text{SEP}}S(\rho||\sigma
).\label{properties}%
\end{equation}
This is monotonic under trace-preserving LOCCs $\Lambda$, i.e.,
$E_{R}[\Lambda(\rho)]\leq E_{R}\left(  \rho\right)  $, and
sub-additive over tensor products, i.e., $E_{R}\left(
\rho\otimes\sigma\right)  \leq E_{R}\left(  \rho\right)
+E_{R}\left(  \sigma\right)  $.

Now consider the secret-key capacity $K$ of a quantum channel
(maximum number of secret bits per channel use which are generated
by adaptive protocols). This is equal to the two-way private
capacity $P_{2}$\ of the channel (maximum number of private bits
per channel use which are deterministically transmitted from Alice
to Bob by means of adaptive protocols) and greater than the
two-way quantum capacity $Q_{2}$ (maximum number of qubits per
channel use which are reliably sent from Alice to Bob by means of
adaptive protocols). We have the following.

\begin{theorem}
[\cite{Stretching}]The secret key capacity of a channel must
satisfy the weak converse upper bound
\begin{equation}
K(\mathcal{E})\leq E_{R}^{\star}\left(  \mathcal{E}\right)  :=\sup
_{\mathcal{L}}\lim_{n\rightarrow\infty}\frac{E_{R}\left(
\rho_{n}\right)
}{n},\label{UBree}%
\end{equation}
where $\mathcal{L}$ is an adaptive protocol for key generation and
$\rho_{n}$ is its $n$-use output.
\end{theorem}

Now we can see that combining the REE bound in Eq.~(\ref{UBree})
with the stretching in Eq.~(\ref{LOCCsim}), and exploiting the
monotonicity and sub-additivity of the REE, we derive the
following.

\begin{theorem}[\cite{Stretching}]\label{singleletterupper} If a channel
$\mathcal{E}$ is $\tau$-stretchable, then its secret-key capacity
is upper bounded by the REE of its resource state $\tau$, i.e.,
\begin{equation}
K(\mathcal{E})\leq E_{R}\left(  \tau\right)  .
\end{equation}
In particular, for a Choi-stretchable channel, we write
\begin{equation}
K(\mathcal{E})\leq E_{R}\left(  \chi_{\mathcal{E}}\right)  ,
\end{equation}
where $\chi_{\mathcal{E}}$ is its Choi matrix.
\end{theorem}

\section{Simulating non-Pauli channels via ``noisy'' teleportation}\label{Expanding}

Whilst we have an extremely simple way of simulating Pauli
channels, i.e., just standard teleportation on a two-qubit mixed
state~\cite{B2,BoBo}, we would like to have a similarly easy way
for simulating non-Pauli channels. Here we show that this is
possible by means of a simple modification of the teleportation
protocol where we also include a classical channel in the CCs from
Alice\ to Bob. This is non-trivial because until now, the only way
to generate non-Pauli channels via DV teleportation is by changing
the dimension of the Hilbert space between the systems $A$ and $B$
of the shared resource of Alice and Bob (e.g., using a
qubit-qutrit resource state, one may simulate an erasure channel).
In the following discussion, we shall limit ourselves to the case
where $\mathcal{E}$ maps qubits to qubits.

Consider a classical channel $\Pi$ from Alice's outcome $k$ for
the Bell measurement to Bob's variable $l$ for the corrective
Pauli unitary $U_{l}$. This is characterized by conditional
probability distribution~\cite{Shannon}
$\left\{  p_{l|k}\right\}  $ such that%
\begin{equation}
p_{l|k}\geq0,~~\sum_{l=0}^{3}p_{l|k}=1,\;\;\;\forall
k\in\{0,1,2,3\}.
\end{equation}
What this means in practical terms is that when Alice obtains the
Bell outcome $k$, rather than Bob performing the corrective
unitary $U_{k}$ with certainty, instead he performs one of the
four unitaries $U_{l}$ with probability $p_{l|k}$. Using such a
noisy teleportation protocol, we prove the following.

\begin{theorem}
\label{FFormula} Consider a teleportation protocol based on a Bell
detection and Pauli correction unitaries but where the resource
state is a generic two-qubit state $\tau$ and the CCs from Alice
to Bob are subject to a classical channel $\Pi$ (``noisy
teleportation''). In this way, we simulate a quantum channel
$\mathcal{E}_{f}$ whose action on the Bloch sphere is described by
\begin{align}
\mathcal{E}_{f}:\left(  x,y,z\right)  \rightarrow( &  f_{10}+f_{11}%
x+f_{12}y+f_{13}z,\nonumber\\
&  f_{20}+f_{21}x+f_{22}y+f_{23}z,\nonumber\\
&  f_{30}+f_{31}x+f_{32}y+f_{33}z)\label{fullform}%
\end{align}
where $f_{ij}$ is given by the formula
$f_{ij}=t_{ji}^{\prime}S_{ij}$, where
\begin{equation}
S_{ij}:=\frac{1}{4}\sum_{k,l=0}^{3}-1^{\delta_{k,0}+\delta_{j,2}+\delta
_{j,0}+\delta_{k,j}+\delta_{i,l}+\delta_{0,l}}p_{l|k},\label{Sdefinition}%
\end{equation}
and $T^{\prime}$ is defined as the \textquotedblleft augmented"
$T$ matrix,
\begin{equation}
t_{ji}^{\prime}=%
\begin{cases}
b_{i} & j=0\\
t_{ji} & j\in\left\{  1,2,3\right\}
\end{cases}
i\in\left\{  1,2,3\right\}  ,
\end{equation}
taking $t_{ji}$ from the $T$ matrix of Eq.(\ref{generaltwo}).
\end{theorem}

By comparing Eq.~(\ref{TRASF}) with Eq.~(\ref{fullform}), we can see immediately that the inclusion of a classical channel
opens up much wider variety of simulated quantum channels. In
fact, we may now have dependence on $x$, $y$ and $z$ in any part
of the transformed Bloch vector, and it is also possible to add
constant terms. This clearly allows us to go well beyond Pauli
channels (a specific class of non-Pauli channels will be discussed
in the next section). Here we may also state the following result
which is a no-go for the simulation of non-Pauli channels when the
noisy teleportation protocol is restricted to Bell diagonal
resource states.

\begin{theorem}\label{nogo}
Using a Bell diagonal resource state, i.e., of the form in
Eq.~(\ref{PauliChoi}), it is only possible to simulate Pauli
channels regardless of the classical channel in place between the
two parties.
\end{theorem}

\noindent\textbf{Proof.} From the structure of $S_{ij}$, we can
see it can only take values in $[-1,1]$. Making use
of~(\ref{fullform}), we see that the action of any channel
generated using resource state~(\ref{PauliChoi}) will be
\begin{equation}
\mathcal{E}:(x,y,z)\rightarrow(t_{11}S_{11}x,t_{22}S_{22}y,t_{33}S_{33}z)~.
\end{equation}
Looking at the structure of the sums $S_{ii}$ for $i\in\{1,2,3\}$
(given in Appendix~\ref{App:sumforms}), we find that for any valid
$p_{l|k}$ term within the sum induces one of four transformations
\begin{align}
\mathcal{E}_{p_{l|k}}:(x,y,z) &  \rightarrow(\phantom{-}t_{11}x,-t_{22}%
y,\phantom{-}t_{33}z)\label{idch}\\
&  \rightarrow(\phantom{-}t_{11}x,\phantom{-}t_{22}y,-t_{33}z)\label{sxch}\\
&  \rightarrow(-t_{11}x,-t_{22}y,-t_{33}z)\label{sych}\\
&  \rightarrow(-t_{11}x,\phantom{-}t_{22}y,\phantom{-}t_{33}z)~,\label{szch}%
\end{align}
which are the four Pauli transformations induced by simulation
over the respective states defined by
\begin{align*}
&  (\phantom{-}t_{11},\phantom{-}t_{22},\phantom{-}t_{33}), &  &
(\phantom{-}t_{11},-t_{22},-t_{33}),\\
&  (-t_{11},\phantom{-}t_{22},-t_{33}), &  &  (-t_{11},-t_{22}%
,\phantom{-}t_{33}),
\end{align*}
with perfect classical communication. We have assumed that $(t_{11}%
,t_{22},t_{33})$ is given by a convex weighting of our four bell
states with some probabilities $p_{i}$, and it is easy to spot
that we may obtain the other three states from the Bell states by
permuting these weights. Since the set $\{\frac{p_{l|k}}{4}\}$
sums to 1, this may also be thought of as a convex
weighting, and thus we may conclude that $(t_{11}S_{11},t_{22}S_{22}%
,t_{33}S_{33})\in\mathcal{T}$, and so induces a Pauli
channel.$~\square$

It is important to understand the difference between
Theorem~\ref{nogo} and Theorem~\ref{Bobothe}.
Theorem~\ref{Bobothe} tells us that an \textit{arbitrary} two
qubit resource state with \textit{perfect} CC from Alice to Bob
may only simulate Pauli channels, whereas Theorem~\ref{nogo}
states that a \textit{Bell diagonal} resource with an
\textit{arbitrary} classical channel for the CC from Alice to Bob
may only simulate Pauli channels. As a result, we have the
following corollary which will drive us in the choice of the
resource state in the next section.

\begin{corollary}\label{coro}
In order to simulate a non-Pauli channel via noisy teleportation,
the resource state $\tau$ of Eq.~(\ref{generaltwo}) must
have $\mathbf{b}\neq0$ or $T$ non-diagonal. This means $\tau$
cannot be the Choi matrix of a Pauli channel.
\end{corollary}

\section{Amplitude damping as a resource for simulating non-Pauli
channels}\label{dampSEC} Following Corollary~\ref{coro}, we will
explore resource states which are non-diagonal in the Bell basis.
A natural choice is to consider the Choi matrix of the amplitude
damping channel. This is the most studied (dimension preserving)
non-Pauli channel. It has the action
\begin{align}
\mathcal{E}_{\gamma}:\ket{0} &  \rightarrow\ket{0},\\
\ket{1} &  \rightarrow\sqrt{\gamma}\ket{0}+\sqrt{1-\gamma}\ket{1},
\end{align}
where $\gamma\in\lbrack0,1]$ is the probability of damping.
Alternatively, on the Bloch sphere, we have
\begin{equation}
\mathcal{E}_{\gamma}:(x,y,z)\rightarrow\left(
\sqrt{1-\gamma}x,\sqrt {1-\gamma}y,\gamma+(1-\gamma)z\right)  .
\end{equation}
The Choi matrix of this channel is
\begin{equation}
\chi_{\gamma}=\left(
\begin{array}
[c]{cccc}%
\frac{1}{2} & 0 & 0 & \frac{\sqrt{1-\gamma}}{2}\\
0 & 0 & 0 & 0\\
0 & 0 & \frac{\gamma}{2} & 0\\
\frac{\sqrt{1-\gamma}}{2} & 0 & 0 & \frac{1-\gamma}{2}%
\end{array}
\right),
\end{equation}
which is a resource state of the form (\ref{generaltwo}), where
the non-zero entries are only
\begin{equation}
b_{3}=\gamma,~t_{11}=\sqrt{1-\gamma},~t_{22}=-\sqrt{1-\gamma},~t_{33}=1-\gamma.
\end{equation}

It is useful to define the \emph{F matrix} of a channel, which
compactly describes the action of the channel on the augmented
Bloch vector $(1,x,y,z)$.

\begin{definition}
\label{FMatrix} A quantum channel
$\mathcal{E}:(x,y,z)\rightarrow(x^{\prime
},y^{\prime},z^{\prime})$ can be described by its F matrix
$F_{\mathcal{E}}$, where
\begin{equation}
\left(
\begin{array}
[c]{c}%
1\\
x^{\prime}\\
y^{\prime}\\
z^{\prime}%
\end{array}
\right)  =F_{\mathcal{E}}\left(
\begin{array}
[c]{c}%
1\\
x\\
y\\
z
\end{array}
\right)  =\left(
\begin{array}
[c]{cccc}%
1 & 0 & 0 & 0\\
f_{10} & f_{11} & f_{12} & f_{13}\\
f_{20} & f_{21} & f_{22} & f_{23}\\
f_{30} & f_{31} & f_{32} & f_{33}%
\end{array}
\right)  \left(
\begin{array}
[c]{c}%
1\\
x\\
y\\
z
\end{array}
\right)  .
\end{equation}

\end{definition}

The F matrix of an amplitude damping channel $\mathcal{E}_{\gamma
}$ is
\begin{equation}
F_{\gamma }=\left(
\begin{array}{cccc}
1 & 0 & 0 & 0 \\
0 & \sqrt{1-\gamma } & 0 & 0 \\
0 & 0 & \sqrt{1-\gamma } & 0 \\
\gamma  & 0 & 0 & 1-\gamma
\end{array}%
\right) .
\end{equation}%
For a Pauli channel $\mathcal{E}:\left( x,y,z\right) \rightarrow
\left(
t_{11}x,-t_{22}y,t_{33}z\right) $, we may set $q_{i}:=t_{ii}$\ and write%
\begin{equation}
F_{P}=\left(
\begin{array}{cccc}
1 & 0 & 0 & 0 \\
0 & q_{1} & 0 & 0 \\
0 & 0 & -q_{2} & 0 \\
0 & 0 & 0 & q_{3}%
\end{array}%
\right) ,
\end{equation}%
with $\mathbf{q}=(q_{1},q_{2},q_{3})$ belonging to the tetrahedron
$\mathcal{T}$ (see Sec.~\ref{ResChange}).

We are now ready to present the first of our two main results,
where we provide the general form of the channel that are
simulable by noisy teleportation over the Choi matrix
$\chi_{\gamma}$ of the amplitude damping channel.

\begin{theorem}\label{mainresult}
All channels that are simulable by noisy teleportation over the
amplitude damping Choi matrix $\chi_\gamma$ can be uniquely
decomposed in the following way
\begin{equation}\label{decomposition}
\mathcal{E}_\text{sim}=\sigma_x^{u}\circ\mathcal{E}_\eta\circ\mathcal{E}_{P}
\end{equation}
where $u=0$ or $1$, $\sigma_x$ is the Pauli unitary
$\sigma_x(\rho)=\sigma_x\rho\sigma_x^\dagger$,
$\mathcal{E}_\eta$ is an amplitude damping channel with parameter $\eta$, and $\mathcal{E}_{P}$ is a Pauli channel with suitable parameters $\mathbf{q}=(q_{1}%
,q_{2},q_{3})$ belonging to the tetrahedron $\mathcal{T}$.
\end{theorem}

\noindent\textbf{Proof.} Making use the formula in Eq.
(\ref{fullform}) we know that any channel
$\mathcal{E}_{\text{sim}}$ simulated with $\chi_\gamma$ will have
$F$ matrix
\begin{equation}
F_{\text{sim}}=\left(
\begin{array}{cccc}
1 & 0 & 0 & 0\\
0 & \sqrt{1-\gamma}S_{11} & 0 & 0\\
0 & 0 & -\sqrt{1-\gamma}S_{22} & 0 \\
\gamma S_{30} & 0 & 0 & (1- \gamma) S_{33}
\end{array}
\right).
\end{equation}
If two channels have identical $F$ matrices, then they are
equivalent. This is because they both enact the same action on an
arbitrary qubit state. Thus we aim to prove the theorem by
equating the above $F$ matrix of a simulated channel with that of our
decomposition defined in Eq.~(\ref{decomposition}). From the $F$ matrices
of $\mathcal{E}_{\eta }$ and $\mathcal{E}_{P}$, we derive that $\mathcal{E}%
_{+}:=\mathcal{E}_{\eta }\circ \mathcal{E}_{P}$ and
$\mathcal{E}_{-}:=\sigma _{x}\circ \mathcal{E}_{\eta }\circ
\mathcal{E}_{P}$ have $F$ matrices
\begin{align}
F_{+}& =\left(
\begin{array}{cccc}
1 & 0 & 0 & 0 \\
0 & \sqrt{1-\eta }q_{1} & 0 & 0 \\
0 & 0 & -\sqrt{1-\eta }q_{2} & 0 \\
\eta  & 0 & 0 & (1-\eta )q_{3}%
\end{array}%
\right) , \\
F_{-}& =\left(
\begin{array}{cccc}
1 & 0 & 0 & 0 \\
0 & \sqrt{1-\eta }q_{1} & 0 & 0 \\
0 & 0 & \sqrt{1-\eta }q_{2} & 0 \\
-\eta  & 0 & 0 & -(1-\eta )q_{3}%
\end{array}%
\right) ,
\end{align}%
where $(q_{1},q_{2},q_{3})\in \mathcal{T}$. Since $\eta\geq 0$,
yet $\gamma S_{30}\in[-\gamma,\gamma]$, we are proposing that
\begin{equation}\label{equality}
F_{\text{sim}}
=\begin{cases}
F_+ &\text{if }S_{30}\geq 0 ,\\
F_- &\text{if }S_{30}\leq 0.
\end{cases}
\end{equation}
We will begin by considering the first case where $S_{30}\geq 0$. Equating the $f_{30}$ components it is clear
that we must set $\eta=\gamma S_{30}$. As $S_{30}\leq 1$ this is a valid $\eta$ value. Rearranging (\ref{equality}) this gives us that
\begin{align}
\left(q_1,q_2,q_3\right)=\bigg(\sqrt{\frac{1-\gamma}{1-\gamma S_{30}}}&S_{11},\nonumber\\
\sqrt{\frac{1-\gamma}{1-\gamma S_{30}}}&S_{22},\nonumber\\
\frac{1-\gamma}{1-\gamma S_{30}}&S_{33}\bigg).\label{qvec}
\end{align}

The vector $\left(S_{11},S_{22},S_{33}\right)$ belongs to the
tetrahedron $\mathcal{T}$, which we prove by showing (in Appendix
\ref{App:proofs})
\begin{align*}
S_{11}+S_{22}+S_{33}&\leq 1\\
S_{11}-S_{22}-S_{33}&\leq 1\\
-S_{11}+S_{22}-S_{33}&\leq 1\\
-S_{11}-S_{22}+S_{33}&\leq 1.
\end{align*}
Moreover, the scaling of this vector seen in equation (\ref{qvec}) simply maps to another point still within the tetrahedron
(also proven in Appendix \ref{App:proofs}). Thus we may conclude, in the case where $S_{30}\geq 0$, that our decomposition is valid and unique,
since equality defines a valid value for $\eta$, and a valid point in $\mathcal{T}$ defining $\mathcal{E}_{P}$ given by Eq. (\ref{qvec}).\\
The proof for the case when $S_{30}\leq 0$ is very similar to the first case, therefore we have included it in Appendix \ref{App:proofs}.
$~\square$

\subsection{Pauli-damping channels}\label{PossSim}

We have shown that all the channels simulable by noisy
teleportation over the resource state $\chi_{\gamma}$ are
necessarily of the form~(\ref{decomposition}). Here we discuss the
converse, i.e., we establish what channels of this form are
simulable, i.e., the region of parameters that are accessible in
the parametrization of Eq.~(\ref{decomposition}). This is the
content of the following theorem.

\begin{theorem}
\label{secondmain} Using noisy teleportation over the amplitude
damping Choi matrix $\chi_{\gamma}$, it is only possible to
simulate channels of the form in
Eq.~(\ref{decomposition}) where $\eta\in\lbrack0,\gamma]$ and $\mathbf{q}=(q_{1}%
,q_{2},q_{3})$ belonging to the convex space bounded by the points
\begin{align}
&\left(   \phantom{-}a\phantom{b}  ,   \pm ab     
,\mp a^2b  \right),\nonumber\\
&\left(   \pm ab    , 
\phantom{-}a\phantom{b}   
,\mp a^2b  \right),\nonumber\\
&\left(   -a\phantom{b}  ,   \pm ab     
,\pm a^2b  \right),\nonumber\\
&\left(   \pm ab   ,   -a\phantom{b}   
,\pm a^2b  \right),
\end{align}
with 
\begin{equation*}
a=\sqrt{\frac{1-\gamma}{1-\eta}},\;b=1-\frac{\eta}{\gamma}.
\end{equation*}
These correspond to the extremal points of the tetrahedron
$\mathcal{T}$ truncated by the two planes $z=\pm b  $, and shrunk by the transformation
\begin{equation}
(x,y,z)\rightarrow\left(
a x,a y,a^2 z\right)  .
\end{equation}
\end{theorem}

This theorem motivates the following definition.

\begin{definition}
We define the Pauli-damping channels as the class of qubit
channels that are simulable by teleporting over amplitude damping
Choi matrix $\chi_{\gamma}$ and using a classical channel $\Pi$
for the CCs. They have a unique decomposition form in
Theorem~\ref{mainresult}, and must satisfy the criteria in
Theorem~\ref{secondmain}.
\end{definition}

\noindent\textbf{Proof.} First we consider $\mathcal{E}_{\eta}$.
Since $\eta=|\gamma S_{30}|$, and $S_{30}$ can take any value in
$[-1,1]$, we can conclude that $\eta\in\lbrack0,\gamma]$. A
slightly trickier question now arises: Given our resource has
parameter $\gamma$, and our amplitude damping channel within the
decomposition has parameter $\eta$, what Pauli channels are
attainable? We know that in our two cases (positivity/negativity
of $S_{30}$), the Pauli channel elements $\mathbf{q}$ are
\begin{align}
\text{case 1} &  \text{:}\nonumber\\
\mathbf{q} &  =\left(
\begin{array}
[c]{c}%
\frac{\sqrt{1-\gamma}}{\sqrt{1-\gamma S_{30}}}S_{11},\\
\phantom{-}\frac{\sqrt{1-\gamma}}{\sqrt{1-\gamma S_{30}}}S_{22},\\
\phantom{-}\frac{1-\gamma}{1-\gamma S_{30}}S_{33}%
\end{array}
\right)  =\left(
\begin{array}
[c]{c}%
\frac{\sqrt{1-\gamma}}{\sqrt{1-|\gamma S_{30}|}}S_{11},\\
\phantom{-}\frac{\sqrt{1-\gamma}}{\sqrt{1-|\gamma S_{30}|}}S_{22},\\
\phantom{-}\frac{1-\gamma}{1-|\gamma S_{30}|}S_{33}%
\end{array}
\right),  \\
\text{case 2} &  \text{:}\nonumber\\
\mathbf{q} &  =\left(
\begin{array}
[c]{c}%
\frac{\sqrt{1-\gamma}}{\sqrt{1+\gamma S_{30}}}S_{11},\\
-\frac{\sqrt{1-\gamma}}{\sqrt{1+\gamma S_{30}}}S_{22},\\
-\frac{1-\gamma}{1+\gamma S_{30}}S_{33}%
\end{array}
\right)  =\left(
\begin{array}
[c]{c}%
\frac{\sqrt{1-\gamma}}{\sqrt{1-|\gamma S_{30}|}}S_{11},\\
-\frac{\sqrt{1-\gamma}}{\sqrt{1-|\gamma S_{30}|}}S_{22},\\
-\frac{1-\gamma}{1-|\gamma S_{30}|}S_{33}%
\end{array}
\right)  .
\end{align}
Since we may prove that both
\begin{equation}
(S_{11},S_{22},S_{33}),(S_{11},-S_{22},-S_{33})\in\mathcal{T},
\end{equation}
(see Lemma~\ref{Sin} in Appendix \ref{App:proofs}), then we can
state with certainty that the class of possible Pauli channels
will be bound by the \textquotedblleft shrunk" tetrahedron
\begin{align}
&  \left(
\phantom{-}\frac{\sqrt{1-\gamma}}{\sqrt{1-\eta}},\phantom{-}\frac
{\sqrt{1-\gamma}}{\sqrt{1-\eta}},-\frac{1-\gamma}{1-\eta}\right)  \nonumber\\
&  \left(  \phantom{-}\frac{\sqrt{1-\gamma}}{\sqrt{1-\eta}},-\frac
{\sqrt{1-\gamma}}{\sqrt{1-\eta}},\phantom{-}\frac{1-\gamma}{1-\eta}\right)
\nonumber\\
&  \left(  -\frac{\sqrt{1-\gamma}}{\sqrt{1-\eta}},\phantom{-}\frac
{\sqrt{1-\gamma}}{\sqrt{1-\eta}},\phantom{-}\frac{1-\gamma}{1-\eta}\right)
\nonumber\\
&  \left(  -\frac{\sqrt{1-\gamma}}{\sqrt{1-\eta}},-\frac{\sqrt{1-\gamma}%
}{\sqrt{1-\eta}},-\frac{1-\gamma}{1-\eta}\right)  .\label{positiveshrink}%
\end{align}
As well as this, we fixed the value of $S_{30}$ when choosing our
$\eta$ value. Since $S_{11},S_{22},S_{33}$ are dependent of the
same variables as $S_{30}$, this places some restrictions of the
values they may take. In order to obtain this, we first use vertex
enumeration~\cite{PANDA} to find all extremal probability
distributions of the space defined by
\begin{align}
\mathcal{P}_{\eta}^{\pm}=\bigg\{p_{l|k}\mid p_{l|k}\geq0, &
\;\;\sum
_{k=0}^{3}p_{l|k}=1,\\
&  \;\;S_{30}=\pm\frac{\eta}{\gamma}\;\;k,l\in\left\{
0,1,2,3\right\} \bigg\},\nonumber
\end{align}
which we will denote $\left\{  Q_{m}^{\pm}\right\}  $. Now we may
consider $(S_{11},S_{22},S_{33}),\,(S_{11},-S_{22},-S_{33})$ as
two linear functions, $\mathcal{S}_{+}$ and $\mathcal{S}_{-}$,
which map
\[
\mathcal{S}_{\pm}:\mathcal{P}_{\eta}^{\pm}\rightarrow\mathcal{T},
\]
Thus for a given probability distribution $\Pi$, we may calculate
this
transformation as%
\begin{equation}
\mathcal{S}_{\pm}(\Pi)=\mathcal{S}_{\pm}\left(
\sum_{m}\lambda_{m}Q_{m}^{\pm }\right)
=\sum_{m}\lambda_{m}\mathcal{S}_{\pm}\left(  Q_{m}^{\pm}\right)  ,
\end{equation}
with $\sum_{m}\lambda_{m}=1,\;\lambda_m\geq 0$. Therefore, we need only consider the
values of $S_{\pm}$ at these extremal probability distributions,
in order to obtain all allowable $S_{ii}$ values. These are easily
calculated, and we obtain that the
eight extremal distributions are%
\begin{align}
\bigg( &  \phantom{-}1 & , &  \pm\left(
1-\frac{\eta}{\gamma}\right)   &  &
,\mp\left(  1-\frac{\eta}{\gamma}\right)  \bigg),\nonumber\\
\bigg( &  \pm\left(  1-\frac{\eta}{\gamma}\right)   & , &
\phantom{-}1 &  &
,\mp\left(  1-\frac{\eta}{\gamma}\right)  \bigg),\nonumber\\
\bigg( &  -1 & , &  \pm\left(  1-\frac{\eta}{\gamma}\right)   &  &
,\pm\left(  1-\frac{\eta}{\gamma}\right)  \bigg),\nonumber\\
\bigg( &  \pm\left(  1-\frac{\eta}{\gamma}\right)   & , &  -1 &  &
,\pm\left(  1-\frac{\eta}{\gamma}\right)  \bigg),
\end{align}
regardless of the case ($S_{30}$ positive or negative). These
points correspond to $\mathcal{T}$, truncated by two planes at
$S_{33}=\pm\left( 1-\frac{\eta}{\gamma}\right)  $.~$\square$

\begin{figure}[h!]
\begin{centering}
\includegraphics[width=8cm]{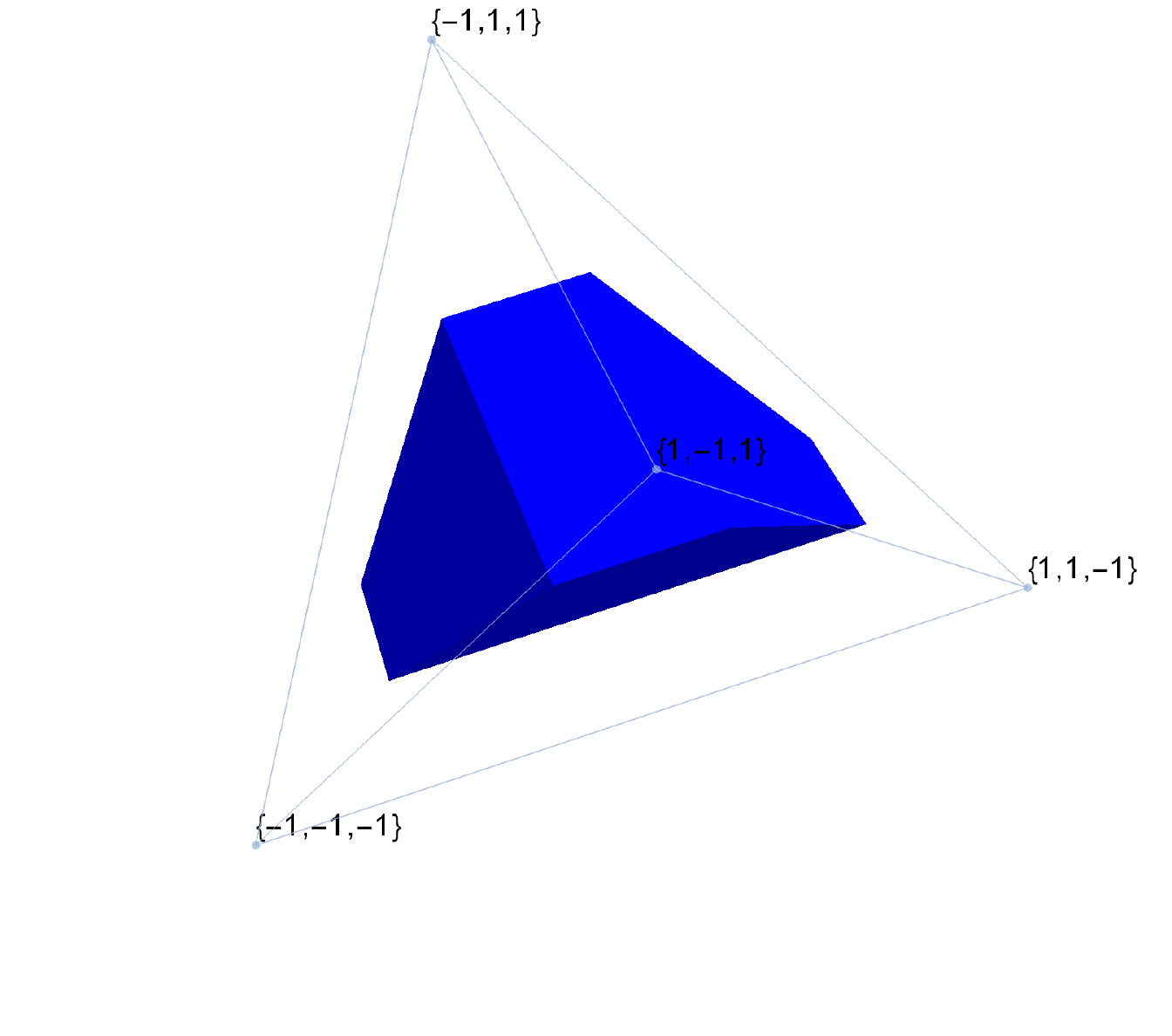}
\vspace{-1cm} \caption{Possible Pauli channels when $S_{30}=0.5$
and $\gamma=0.6$, including the shrinking effect of
Eq.~(\ref{positiveshrink}). The hollow tetrahedron is
$\mathcal{T}$ characterizing all Pauli channels, whilst the shaded
region is the allowable values of $\mathbf{q}$ bounded by $
(\;\;\;\frac{2}{\sqrt{7}} ,\pm\frac{1}{\sqrt{7}} ,
\mp\frac{2}{7}), (\pm\frac{1}{\sqrt{7}} ,\;\;\;\frac{2}{\sqrt{7}}
, \mp\frac{2}{7}),
(-\frac{2}{\sqrt{7}} ,\pm\frac{1}{\sqrt{7}} , \pm\frac{2}{7}),\\
(\pm\frac{1}{\sqrt{7}} ,-\frac{2}{\sqrt{7}} , \pm\frac{2}{7})
$ for these particular values.
}\label{Truly}
\end{centering}
\end{figure}

An immediate consequence of this theorem is that we cannot
simulate the amplitude damping channel $\mathcal{E}_{\gamma}$
using its Choi matrix
$\chi_{\gamma}$.\ In fact, this would require $\eta=\gamma$ and $\mathcal{E}%
_{P}=\mathbb{I}$, corresponding to $\mathbf{q}=(1,-1,1)$. However,
when $\eta=\gamma$ our possible Pauli channels are limited from
both above and below by the same plane,
$q_{3}=\pm\frac{1-\gamma}{1-\gamma}\left(
1-\frac{\gamma}{\gamma}\right)  =0$, and thus this is impossible.
Therefore the amplitude damping channel is not Choi-stretchable
even with the noisy teleportation protocol. The only exceptions to
this are the special cases where $\gamma=0$, which is simply the
identity channel, and when $\gamma=1$, which sends all qubit
states deterministically to $\ket{0}$. This can be decomposed into
the completely depolarizing channel $\mathcal{E}_{D}$ with
$\mathbf{q}=\mathbf{0}$, which sends all states to the maximally
mixed state $\frac{\mathbb{I}}{2}$, followed by itself, to fit our
decomposition (see Fig.~\ref{Truly}).

\section{Properties and capacities of Pauli-damping
channels}\label{SECprop}

Now that we have shown what channels can be simulated, we study
some of the properties of these channels. First of all, we
quantify how distinguishable they are from their closest Pauli
equivalent. It turns out that the decomposition in
Theorem~\ref{mainresult} provides a simple answer to this problem:
the distance is simply $\eta$.

\subsection{Distance in trace norm}

The trace norm  distance between two quantum channels
$\mathcal{E}_{1}$ and $\mathcal{E}_{2}$ can be defined as
\begin{equation}
||\mathcal{E}_{1}-\mathcal{E}_{2}||_{1}:=\sup_{\rho}||\mathcal{E}_{1}\left(
\rho\right)  -\mathcal{E}_{2}\left(  \rho\right)  ||_{1}%
\end{equation}
where $||\sigma||_{1}=$\textrm{Tr}$\sqrt{\sigma\sigma^{\dagger}}$.
For Hermitian matrices, this is equivalent to the sum of the
absolute values of the eigenvalues of $\sigma$. We then state the
following.

\begin{proposition}\label{postrace}
Given a decomposition
$\mathcal{E}_{\text{sim}}=\sigma_{x}^u\circ\mathcal{E}_{\eta}\circ\mathcal{E}_{P}$
characterized by $\eta$ and $(q_{1}%
,q_{2},q_{3})$ respectively, then the trace norm between $\mathcal{E}%
_{\text{sim}}$ and the closest Pauli channel
$\mathcal{E}_{\text{cl}}$ is
simply $\eta$. Moreover, the closest Pauli channel has $(f_{11},f_{22}
,f_{33})=\begin{cases}
\left(
\sqrt{1-\eta}q_{1},-\sqrt{1-\eta}q_{2},\phantom{-}(1-\eta)q_{3}\right) & \mbox{ for } u=0,\\
\left(
\sqrt{1-\eta}q_{1},\phantom{-}\sqrt{1-\eta }q_{2},-(1-\eta)q_{3}\right)& \mbox{ for } u=1.
\end{cases}$
\end{proposition}

\noindent\textbf{Proof.} For qubits, the trace norm between
two states is simply the Euclidean distance between their Bloch
vectors. Therefore we have a very natural way to find the trace
norm between two-qubit channels. When $u=0$, the Bloch vector of a state under $\mathcal{E}_{\text{sim}}$ is $\mathbf{r}_{\text{sim}}=(\sqrt{1-\eta}q_{1}x,-\sqrt{1-\eta
}q_{2}y,(1-\eta)q_{3}z+\eta)$, whilst under an arbitrary Pauli channel it is $\mathbf{r}_P=(c_1x,-c_2y,c_3z)$. Thus the problem we need to solve is
\begin{align}
&  \min_{\left(  c_{1},c_{2},c_{3}\right)  \in\mathcal{T}}\;\;\max
_{x,y,z:x^{2}+y^{2}+z^{2}\leq1}\nonumber\\
&  \left(  (\sqrt{1-\eta}q_{1}-c_{1})x\right)  ^{2}+\left(
-(\sqrt{1-\eta
}q_{2}-c_{2})y\right)  ^{2}\nonumber\\
+ &  \Big(((1-\eta)q_{3}-c_{3})z+\eta\Big)^{2},
\end{align}
which is the square of the trace norm. Let us first look at the
final term $\left(  \left(  (1-\eta)q_{3}-c_{3}\right)
z+\eta\right)  ^{2}$. Given our maximum occurs for some fixed
$|z|$ value, we have that the value of this term will be
\begin{align}
\max\bigg\{ &  \big(\left(  (1-\eta)q_{3}-c_{3}\right)  |z|+\eta
\big)^{2},\nonumber\\
&  \big(-\left(  (1-\eta)q_{3}-c_{3}\right)  |z|+\eta\big)^{2}%
\bigg\}\nonumber\\
= &  \big(|(1-\eta)q_{3}-c_{3}||z|+\eta\big)^{2}.
\end{align}
Clearly this is minimized when $c_{3}=\left(  1-\eta\right)
q_{3}$, and has value $\eta^{2}$.

The remaining two parts of the equation are simpler. Clearly we
want to set
\[
c_{1}=\sqrt{1-\eta}q_{1},~c_{2}=\sqrt{1-\eta}q_{2},
\]
to make these parts disappear, regardless of the values of $x$ and
$y$. Thus we obtain our closest Pauli channel to be
\begin{equation}
(x,y,z)\rightarrow(\sqrt{1-\eta}q_{1}x,-\sqrt{1-\eta}q_{2}y,(1-\eta)q_{3}z).
\end{equation}
We can be sure that this channel is Pauli as a consequence of Lemma~\ref{shrinking}
in Appendix~\ref{App:proofs}.\\
For the case when $u=1$, the proof is very similar, and given in Appendix \ref{App:proofs}.$~\square$

\subsection{Distance in diamond norm}
It is not wise to use the trace norm as a measure for the
distinguishability of channels, since it has been shown we can do
it better in general by sending part of an entangled state through
the channel~\cite{Ent1,Ent2,Ent3,Ent4,Ent5,Ent6}. With this in
mind, we look to an alternative distance. The diamond norm
distance $||\mathcal{E}_{1}-\mathcal{E}_{2}||_{\diamond}$ is
defined as:
\begin{equation}
||\mathcal{E}_{1}-\mathcal{E}_{2}||_{\diamond}:=\sup_{\rho\in\kappa
\otimes\mathcal{H}}||\mathbb{I_{\kappa}}\otimes\mathcal{E}_{1}\left(
\rho\right)  -\mathbb{I_{\kappa}}\otimes\mathcal{E}_{2}\left(
\rho\right) ||_{1},
\end{equation}
where $\mathrm{\kappa}$ in an ancillary Hilbert space to the one
acted upon by
$\mathcal{E}$, $\mathcal{H}$. In general, one has $||\mathcal{E}%
_{1}-\mathcal{E}_{2}||_{\diamond}\geq||\mathcal{E}_{1}-\mathcal{E}_{2}||_{1}$.
Also we know that the diamond norm can be achieved with an
ancillary Hilbert space $\kappa$ with
$\text{dim}\;\kappa=\text{dim}\;\mathcal{H}~$\cite{Fano}.
Therefore, we need only consider a 1 qubit ancillary space in our
case and state the following.

\begin{proposition}\label{posdiamond}
For a channel
$\mathcal{E}_\text{sim}=\sigma_{x}^u\circ\mathcal{E}_{\eta}\circ\mathcal{E}_{P}$,
the closest Pauli channel under the diamond norm is the same as
under the trace norm, given in Proposition \ref{postrace}, and the diamond norm  distance is equal to
$\eta$.
\end{proposition}

\noindent\textbf{Sketch Proof.} (Full proof in appendix
\ref{App:proofs}). First off, we know that
\begin{equation}
\min_{\mathcal{E}_2\in\text{Pauli}}||\mathcal{E}_{\text{sim}}-\mathcal{E}_2||_\diamond\leq||\mathcal{E}_{\text{sim}}-\mathcal{E}_{\text{cl}}||_\diamond.
\end{equation}
In order to find the diamond norm between $\mathcal{E}_\text{sim}$ and $\mathcal{E}_{\text{cl}}$, we look at
\begin{equation}
||\mathbb{I}_2\otimes\mathcal{E}_\text{sim}\left(\rho\right)-
\mathbb{I}_2\otimes\mathcal{E}_\text{cl}\left(\rho\right)||_1
\end{equation}
for an arbitrary 2 qubit state $\rho$. We find the absolute sum of
eigenvalues for
$\mathbb{I}_2\otimes\mathcal{E}_\text{sim}\left(\rho\right)-
\mathbb{I}_2\otimes\mathcal{E}_\text{cl}\left(\rho\right)$ to be
independent of $\rho$ and equal to $\eta$. Thus we can conclude
that
\begin{equation}
||\mathcal{E}_{\text{sim}}-\mathcal{E}_{\text{cl}}||_\diamond=\eta=||\mathcal{E}_{\text{sim}}-\mathcal{E}_{\text{cl}}||_1.
\end{equation}
Using this, suppose there exists a channel $\mathcal{E}'$ with a strictly smaller diamond norm than our closest channel. Then we have the chain of inequalities
\begin{equation}\label{chain}
||\mathcal{E}_\text{sim}-\mathcal{E}'||_1\leq||\mathcal{E}_\text{sim}-\mathcal{E}'||_\diamond<||\mathcal{E}_\text{sim}-\mathcal{E}_\text{cl}||_\diamond=||\mathcal{E}_\text{sim}-\mathcal{E}_\text{cl}||_1
\end{equation}
leading to a contradiction, since we know the closest channel under trace norm
to be $\mathcal{E}_\text{cl}$. Thus we are forced to conclude that the diamond norm
is smallest between $\mathcal{E}_\text{sim}$ and $\mathcal{E}_\text{cl}$, with distance $\eta$. $\square$\\

The consequence of this result is that we have a natural measure of the generalization allowed by the introduction of classical channels. Given a resource state $\chi_\gamma$, we know that we will be able to simulate channels $||\cdot||_\diamond=\gamma$ distinct from the set of Pauli channels, since that is the largest allowable value $\eta$ may take.

\subsection{Upper bound for the two-way private capacity}

Now that we have characterized the class of Pauli-damping
channels, we are interested in their quantum and private
communication capacities. As explained in the introduction, the
two-way assisted capacities are in general hard to calculate. Yet
because we have shown that these channels can be simulated with an
LOCC protocol (noisy teleportation) over a pre-shared resource
(the amplitude damping Choi matrix $\chi_{\gamma}$), we may use
teleportation stretching and Theorem~\ref{singleletterupper} to
upper-bound their two-way quantum ($Q_{2}$) and private capacities
($P_{2}=K$). In fact, for an arbitrary Pauli-damping channel
$\mathcal{E}$ with resource state $\chi_{\gamma}$, we may compute
the upper bound (weak converse)
\begin{gather}
Q_{2}(\mathcal{E})\leq P_{2}(\mathcal{E})=K(\mathcal{E})\leq
E_{R}\left(
\chi_{\gamma}\right)  \nonumber\\
\leq\frac{1}{2}-\frac{1-\gamma}{2}\log_{2}\left(
\frac{1-\gamma}{2}\right)
+\frac{2-\gamma}{2}\log_{2}\left(  \frac{2-\gamma}{2}\right)  .\label{YBsss}%
\end{gather}

\begin{figure}[h!]
\begin{centering}
\includegraphics[width=8cm,height=5.3cm]{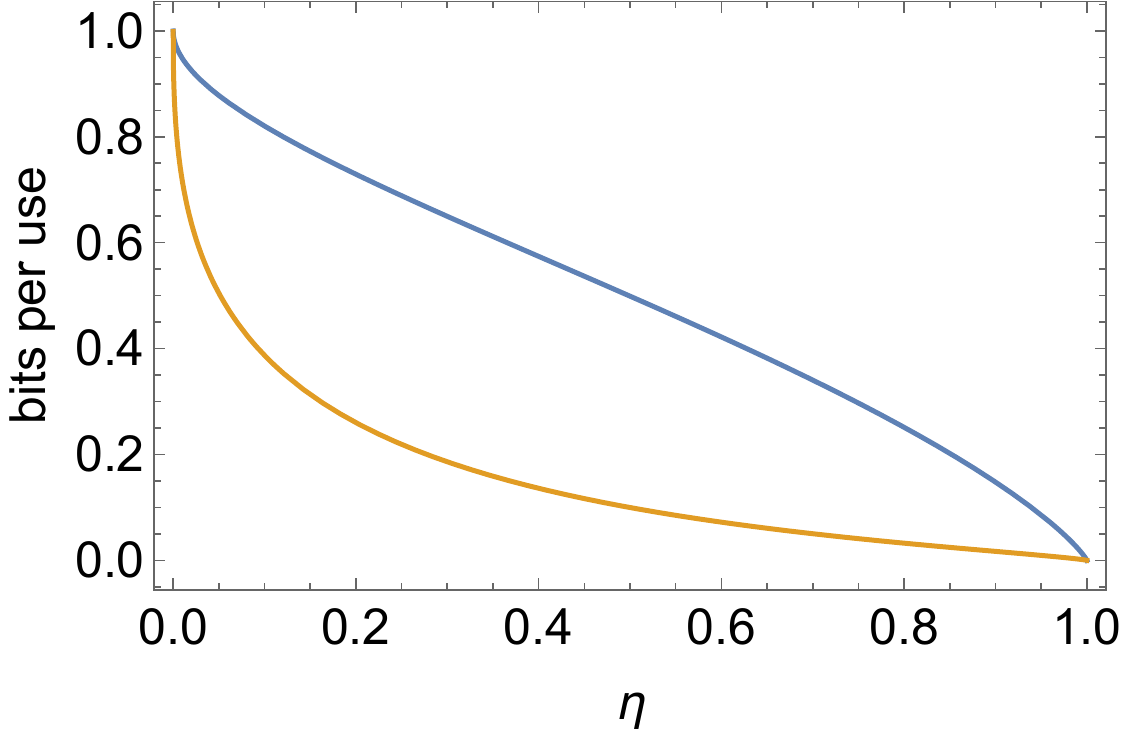}
\vspace{-.3cm} \caption{Upper and lower bounds for the two-way
private capacity $P_2$ and the two-way quantum capacity $Q_2$ of
the squared channel $\mathcal{E}_{\text{sq}}$, in terms of its
parameter $\eta$ which is the square of the amplitude damping
parameter $\gamma$ associated with its resource
state.}\label{Lower}
\end{centering}
\end{figure}

Within the Pauli-damping class, let us analyze the
\textquotedblleft squared\textquotedblright\ channel
$\mathcal{E}_{\text{sq}}$ with its F matrix being given by
\begin{equation}
F_{\text{sq}}=\left(
\begin{array}
[c]{cccc}%
1 & 0 & 0 & 0\\
0 & \sqrt{1-\gamma}\left(  1-\frac{\gamma}{2}\right)   & 0 & 0\\
0 & 0 & \sqrt{1-\gamma}\left(  1-\frac{\gamma}{2}\right)   & 0\\
\gamma^{2} & 0 & 0 & (1-\gamma)^{2}%
\end{array}
\right)  .
\end{equation}
The decomposition of this channel into the
form~(\ref{decomposition}) of Theorem~\ref{mainresult} is $u=0$,
$\eta=\gamma^{2}$, and
\begin{equation}
\mathbf{q}=\left(  \frac{\left(  1-\frac{\gamma}{2}\right)  }{\sqrt{1+\gamma}%
},-\frac{\left(  1-\frac{\gamma}{2}\right)
}{\sqrt{1+\gamma}},\frac{1-\gamma }{1+\gamma}\right),
\end{equation}
where $\gamma$ is the damping parameter of the resource state. Its
two-way quantum and private capacities are upper bounded by using
Eq.~(\ref{YBsss}) and lower bounded by optimizing the coherent
information of the channel. The results are shown in
Fig.~\ref{Lower}.

\section{Conclusions}\label{SECconclu}
In this paper we have studied a particular design for the LOCC
simulation of quantum channels. This design is based on a modified
teleportation protocol where not only the resource state is
generally mixed (instead of maximally entangled) but also the
classical communication channel between the parties is noisy,
i.e., affected by a classical channel. The latter feature allows
us to simulate family of quantum channels, much larger than the
Pauli class, for which we have provided a characterization in
Theorem \ref{FFormula}.

Starting from the Choi matrix of an amplitude damping channel as a
resource state for the noisy teleportation protocol, we can easily
simulate non-Pauli channels. In particular, we have introduced a
new class of simulable channels, that we have called Pauli-damping
channels. Their distance from the set of Pauli channels can be
quantified in terms of the diamond norm and turns out to be easily
related with the damping probability associated with the
generating Choi matrix. For these Pauli-damping channels we have
then used the method of teleportation stretching to derive upper
bounds for their two-way quantum and private capacities.

In conclusion, our results are useful to shed new light in the
area of channel simulation with direct implications for quantum
and private communication with qubit systems. Further developments
may include the study of Pauli-damping channels in the context of
adaptive quantum metrology \cite{Metro}, or in the setting of
secure quantum networks \cite{networkPIRS}.

\textbf{Acknowledgments}. This work has been supported by the
EPSRC via the `UK Quantum Communications Hub' (EP/M013472/1). T.C. and S.P.
would like to thank discussions with R. Laurenza and C. Ottaviani.  L.H. would like to
thank the ERASMUS program who allowed him to visit the University of
York, where this work has been carried out. T.C. acknowledges
funding from a White Rose Scholarship. L.B. has received
funding for this research from the European Research Council
under the European Union Seventh Framework Programme
(FP7/2007-2013)/ERC Grant Agreement No. 308253 PACOMANEDIA.


\onecolumngrid
\appendix

\section{Proof of Theorem \ref{WHtheorem}}\label{App:proof1}

Let us suppose that the channel $\mathcal{E}(\rho_C)$ is Choi-stretchable via a WH-teleportation protocol. This means that
\begin{align}
\mathcal{E}(\rho_C)&=\Gamma_{\chi_\mathcal{E}}(\rho_C)=\sum_{g\in
G}V_B^\dagger(g)\mathrm{Tr}_{CA}\left(E_{CA}(g)(\rho_C\otimes
\chi_\mathcal{E})\right)V_B(g).
\end{align}
Now consider $V_B^\dagger(h)\mathcal{E}\left(U_C(h)\rho_C
U_C^\dagger(h)\right)V_B(h)$ which expands out to
\begin{align}
&V_B^\dagger(h)\left(\sum_{g\in G}V_B^\dagger(g)\mathrm{Tr}_{CA}\left(E_{CA}(g)(U_C(h)\rho_C U_C^\dagger(h)\otimes \chi_\mathcal{E})\right)V_B(g)\right)V_B(h)\\
=&\sum_{g\in
G}V_B^\dagger(h)V_B^\dagger(g)\mathrm{Tr}_{CA}\left(E_{CA}(g)(U_C(h)\rho_C
U_C^\dagger(h)\otimes
\chi_\mathcal{E})\right)V_B(g)V_B(h).\label{A4}
\end{align}
Since $\left\{V_B(g)\right\}$ is a representation of the WH-group,
we may use $V_B(g)V_B(h)=e^{i\phi(g,h)}V_B(gh)$,\\ where
$e^{i\phi(g,h)}$ is some overall phase.
\begin{align}
\mathrm{Eq.~(\ref{A4})}=&\sum_{g\in G}V_B^{\dagger}(gh)\mathrm{Tr}_{CA}\left( U_C^\dagger(h)E_{CA}(g)U_C(h)(\rho_C \otimes \chi_\mathcal{E})\right)V_B(gh)\\
=&\sum_{g\in G}V_B^{\dagger}(gh)\mathrm{Tr}_{CA}\left( U_C^\dagger(h)U_C^\dagger(g)\ket{\Phi}\bra{\Phi}U_C(g)U_C(h)(\rho_C \otimes \chi_\mathcal{E})\right)V_B(gh)\\
=&\sum_{g\in G}V_B^{\dagger}(gh)\mathrm{Tr}_{CA}\left( U_C^\dagger(gh)\ket{\Phi}\bra{\Phi}U_C(gh)(\rho_C \otimes \chi_\mathcal{E})\right)V_B(gh)\\
=&\sum_{g\in
G}V_B^{\dagger}(gh)\mathrm{Tr}_{CA}\left(E_{CA}(gh)(\rho_C \otimes
\chi_\mathcal{E})\right)V_B(gh).\label{A8}
\end{align}
Now we may use the group property that $gG=G$, for any $g\in G$
\begin{align}
\mathrm{Eq.~(\ref{A8})}=&\sum_{g'\in G}V_B^{\dagger}(g')\mathrm{Tr}_{CA}\left(E_{CA}(g')(\rho_C \otimes \chi_\mathcal{E})\right)V_B(g')\phantom{\mathrm{Tr}_{CA}\left(E_{CA}(g')(\rho_C \otimes \chi_\mathcal{E})\right)}\\
=&\Gamma_{\chi_\mathcal{E}}(\rho_C)=\mathcal{E}(\rho_C).
\end{align}
We can therefore conclude that Choi-stretchable channels via
WH-teleportation are teleportation covariant. $\square$

We could also consider a more general case, where we have a
channel in the form seen in Eq. (\ref{WHform}), but without the
group representation structure. However, we would not expect this
to be covariant, since Eq. (\ref{covdef}) forces
\begin{align*}
V(gh)\mathcal{E}(\rho)V^\dagger(gh)&=\mathcal{E}(U(gh)\rho U^\dagger(gh))\\
&=\mathcal{E}(U(g)U(h)\rho U^\dagger(h)U^\dagger(g))\\
&=V(g)\mathcal{E}(U(h)\rho U^\dagger(h))V^\dagger(g)\\
&=V(g)V(h)\mathcal{E}(\rho)V^\dagger(h)V^\dagger(g).
\end{align*}

\section{The form of $S_{ij}$}\label{App:sumforms}
In this section, we present the 12 possible forms for $S_{ij}$ in a concise way. We present a $3\times4$ matrix, $\boldsymbol{\mathcal{S}}$ of $4\times4$ matrices. The rows of $\boldsymbol{\mathcal{S}}$ correspond to $i=1,2,3$ respectively, and the columns to $j=0,1,2,3$. Given $i,j$\textsuperscript{th} element $\boldsymbol{\mathcal{S}}_{ij}$, $S_{ij}$ can be obtained by the sum $\sum_{k=0,l=0}^{3,3}(\boldsymbol{\mathcal{S}}_{ij})_{k,l}\;p_{l|k}$ - note we are counting the rows and columns of $\boldsymbol{\mathcal{S}}_{ij}$ from $0$.
\[\boldsymbol{\mathcal{S}}=\left(
\begin{array}{cccc}
 \left(
\begin{array}{cccc}
 1 & 1 & -1 & -1 \\
 1 & 1 & -1 & -1 \\
 1 & 1 & -1 & -1 \\
 1 & 1 & -1 & -1 \\
\end{array}
\right) & \left(
\begin{array}{cccc}
 1 & 1 & -1 & -1 \\
 1 & 1 & -1 & -1 \\
 -1 & -1 & 1 & 1 \\
 -1 & -1 & 1 & 1 \\
\end{array}
\right) & \left(
\begin{array}{cccc}
 -1 & -1 & 1 & 1 \\
 1 & 1 & -1 & -1 \\
 -1 & -1 & 1 & 1 \\
 1 & 1 & -1 & -1 \\
\end{array}
\right) & \left(
\begin{array}{cccc}
 1 & 1 & -1 & -1 \\
 -1 & -1 & 1 & 1 \\
 -1 & -1 & 1 & 1 \\
 1 & 1 & -1 & -1 \\
\end{array}
\right) \\[1cm]
 \left(
\begin{array}{cccc}
 1 & -1 & 1 & -1 \\
 1 & -1 & 1 & -1 \\
 1 & -1 & 1 & -1 \\
 1 & -1 & 1 & -1 \\
\end{array}
\right) & \left(
\begin{array}{cccc}
 1 & -1 & 1 & -1 \\
 1 & -1 & 1 & -1 \\
 -1 & 1 & -1 & 1 \\
 -1 & 1 & -1 & 1 \\
\end{array}
\right) & \left(
\begin{array}{cccc}
 -1 & 1 & -1 & 1 \\
 1 & -1 & 1 & -1 \\
 -1 & 1 & -1 & 1 \\
 1 & -1 & 1 & -1 \\
\end{array}
\right) & \left(
\begin{array}{cccc}
 1 & -1 & 1 & -1 \\
 -1 & 1 & -1 & 1 \\
 -1 & 1 & -1 & 1 \\
 1 & -1 & 1 & -1 \\
\end{array}
\right) \\[1cm]
 \left(
\begin{array}{cccc}
 1 & -1 & -1 & 1 \\
 1 & -1 & -1 & 1 \\
 1 & -1 & -1 & 1 \\
 1 & -1 & -1 & 1 \\
\end{array}
\right) & \left(
\begin{array}{cccc}
 1 & -1 & -1 & 1 \\
 1 & -1 & -1 & 1 \\
 -1 & 1 & 1 & -1 \\
 -1 & 1 & 1 & -1 \\
\end{array}
\right) & \left(
\begin{array}{cccc}
 -1 & 1 & 1 & -1 \\
 1 & -1 & -1 & 1 \\
 -1 & 1 & 1 & -1 \\
 1 & -1 & -1 & 1 \\
\end{array}
\right) & \left(
\begin{array}{cccc}
 1 & -1 & -1 & 1 \\
 -1 & 1 & 1 & -1 \\
 -1 & 1 & 1 & -1 \\
 1 & -1 & -1 & 1 \\
\end{array}
\right) \\[1cm]
\end{array}
\right)\]

\section{Proofs}\label{App:proofs}

\begin{lemma}\label{rescalingsize}
$\frac{1-\gamma}{1-\gamma S_{30}}\in[0,1]$
\end{lemma}
\noindent\textbf{Proof.}
Remember that $\gamma\in[0,1]$. Therefore $1-\gamma\in[0,1]$ also.\\
Now, $S_{30}\in[0,1]$ in the first case. Thus,
\begin{align*}
\gamma S_{30}&\leq \gamma\\
\Rightarrow-\gamma S_{30}&\geq -\gamma\\
\Rightarrow1-\gamma S_{30}&\geq 1-\gamma\\
\Rightarrow 1&\geq \frac{1-\gamma}{1-\gamma S_{30}}
\end{align*}
remembering $1-\gamma S_{30}\geq 0$.
$\square$\\
\begin{corollary}
$\sqrt{\frac{1-\gamma}{1-\gamma S_{30}}}\in[0,1]$.
\end{corollary}
\begin{lemma}\label{shrinking}
If a point $(x,y,z)$ belongs to the tetrahedron defined by $\mathcal{T}$, then so too does the point $(\sqrt{\alpha}x,\sqrt{\alpha}y,\alpha z)$,where $\alpha\in[0,1]$
\end{lemma}
\noindent\textbf{Proof.}
Since any point in the tetrahedron can be expressed as a convex combination of the four extremal points in $\mathcal{T}$, it is sufficient to show that the four points,
\begin{align}\label{rescaled}
(\phantom{-}\sqrt{\alpha},-\sqrt{\alpha},\phantom{-}\alpha),&(\phantom{-}\sqrt{\alpha},\phantom{-}\sqrt{\alpha},-\alpha),\\
(-\sqrt{\alpha},\phantom{-}\sqrt{\alpha},\phantom{-}\alpha),&(-\sqrt{\alpha},-\sqrt{\alpha},-\alpha),\nonumber
\end{align}
belong to the tetrahedron (i.e. are themselves a convex combination of the four extremal points), and thus any rescaled tetrahedron point also still remains with the full tetrahedron.\\
Expressing any point as
\begin{align*}
(x,y,z)&=p_0(\phantom{-}1,-1,\phantom{-}1)+p_1(\phantom{-}1,\phantom{-}1,-1)\\&+p_2(-1,-1,-1)+p_3(-1,\phantom{-}1,\phantom{-}1),\\&p_0+p_1+p_2+p_3=1,\;\; p_i\geq 0
\end{align*}
then we can achieve the points in Eq. (\ref{rescaled})
\begin{equation}
\begin{array}{ccccc}
\text{Point} & p_0 & p_1 & p_2 & p_3\\
(\sqrt{\alpha},-\sqrt{\alpha},\alpha)& \frac{(1+\sqrt{\alpha})^2}{4} & \frac{1-\alpha}{4}& \frac{1-\alpha}{4} & \frac{(1-\sqrt{\alpha})^2}{4} \\
(\sqrt{\alpha},\sqrt{\alpha},-\alpha)& \frac{1-\alpha}{4} & \frac{(1-\sqrt{\alpha})^2}{4}& \frac{(1+\sqrt{\alpha})^2}{4} & \frac{1-\alpha}{4} \\
(-\sqrt{\alpha},\sqrt{\alpha},\alpha)&  \frac{(1-\sqrt{\alpha})^2}{4} & \frac{1-\alpha}{4}& \frac{1-\alpha}{4} & \frac{(1+\sqrt{\alpha})^2}{4} \\
(-\sqrt{\alpha},-\sqrt{\alpha},-\alpha)&  \frac{1-\alpha}{4} & \frac{(1+\sqrt{\alpha})^2}{4}& \frac{(1-\sqrt{\alpha})^2}{4} & \frac{1-\alpha}{4} .\\
\end{array}
\end{equation}
The normalization and positivity conditions are easy to verify. $\square$\\
\begin{corollary}
If $\left(S_{11},S_{22},S_{33}\right)$ belongs to tetrahedron $\mathcal{T}$, then so too does
\begin{align}
&\left(\sqrt{\frac{1-\gamma}{1-\gamma S_{30}}}S_{11},\sqrt{\frac{1-\gamma}{1-\gamma S_{30}}}S_{22},\frac{1-\gamma}{1-\gamma S_{30}}S_{33}\right)\\=\;\;&(q_1,q_2,q_3)\nonumber
\end{align}
\end{corollary}
\noindent\textbf{Proof.} We can simply set
$\alpha=\frac{1-\gamma}{1-\gamma S_{30}}$, and apply Lemma
\ref{shrinking}. $\square$
\begin{lemma}\label{Sin}
For all classical channels $\Pi$, as defined in our noisy
teleportation protocol, we have that $(S_{11},S_{22},S_{33})$
belongs to the tetrahedron $\mathcal{T}$.
\end{lemma}
\noindent\textbf{Proof.} An alternative way to define
$\mathcal{T}$ is by four inequalities which are satisfied by all
points within the tetrahedron, namely
\begin{align*}
x+y+z&\leq 1\\
x-y-z&\leq 1\\
-x+y-z&\leq 1\\
-x-y+z&\leq 1
\end{align*}
We have already seen these used in Section \ref{ResChange}.
Testing these with $S_{11},S_{22}$ and $S_{33}$ we find
\begin{align*}
S_{11}+S_{22}+S_{33}&=1-(p_{02}+p_{13}+p_{20}+p_{31})\leq 1\\
S_{11}-S_{22}-S_{33}&=1-(p_{03}+p_{12}+p_{21}+p_{30})\leq 1\\
-S_{11}+S_{22}-S_{33}&=1-(p_{00}+p_{11}+p_{22}+p_{33})\leq 1\\
-S_{11}-S_{22}+S_{33}&=1-(p_{01}+p_{10}+p_{23}+p_{32})\leq 1.
\end{align*}
From this, we can conclude that all $(S_{11},S_{22},S_{33})$ possible belong to the tetrahedron. This immediately gives that, in the case where $S_{30}\geq 0$ our decomposition is a valid one. $\square$\\

\noindent\textbf{Proof of Theorem \ref{mainresult} for $S_{30}\leq 0$.}\\
We have already proven this result to be true for $S_{30}\geq 0$ in the main body of the text. We also need to consider our second case, $S_{30}\leq 0$. Here we set $\eta=-\gamma S_{30}$. This time we obtain

\begin{align}
(q_1,q_2,q_3)=\bigg(\frac{\sqrt{1-\gamma}}{\sqrt{1+\gamma S_{30}}}&S_{11},\nonumber\\
-\frac{\sqrt{1-\gamma}}{\sqrt{1+\gamma S_{30}}}&S_{22},\nonumber\\
-\frac{1-\gamma}{1+\gamma S_{30}}&S_{33}\bigg)
\end{align}
Except for the fact our scaling factor is now
$\frac{1-\gamma}{1+\gamma S_{30}}$, we have a very similar
situation to our first case, except now we need to prove that
$(S_{11},-S_{22},-S_{33})$  is in the tetrahedron, in order for
our decomposition to be valid for our second scenario. If we look
at the four inequalities that we need to satisfy, we find that
\begin{align*}
S_{11}+(-S_{22})+(-S_{33})&=S_{11}-S_{22}-S_{33}\\
S_{11}-(-S_{22})-(-S_{33})&=S_{11}+S_{22}+S_{33}\\
-S_{11}+(-S_{22})-(-S_{33})&=-S_{11}-S_{22}+S_{33}\\
-S_{11}-(-S_{22})+(-S_{33})&=-S_{11}+S_{22}-S_{33}.
\end{align*}
which we  already know satisfy our tetrahedron inequalities. Thus
we have proved that our decomposition is valid too for cases where
$S_{30}\leq0$, and so is true for all channels simulable with
$\chi_{\gamma}$ as a resource.$\square$\\

\noindent\textbf{Proof of Proposition \ref{postrace} with $u=1$.}\\
In this case, we have to contend with the
sign change enacted by $\sigma_{x}$; however the proof is
similar. This time, we aim to solve
\begin{align}
&  \min_{\left(  c_{1},c_{2},c_{3}\right)  \in\mathcal{T}}\;\;\max
_{x,y,z:x^{2}+y^{2}+z^{2}\leq1}\nonumber\\
&  \left(  (\sqrt{1-\eta}q_{1}-c_{1})x\right)  ^{2}+\left(
(\sqrt{1-\eta
}q_{2}+c_{2})y\right)  ^{2}\nonumber\\
+ &  \Big((-(1-\eta)q_{3}-c_{3})z-\eta\Big)^{2}.
\end{align}
Again, we begin by looking at the final part of the sum. For a
fixed value of
$|z|$, this term will be%
\begin{align}
\max\bigg\{ &  \big(\left(  -(1-\eta)q_{3}-c_{3}\right)  |z|-\eta
\big)^{2},\nonumber\\
&  \big(\left(  (1-\eta)q_{3}+c_{3}\right)  |z|-\eta\big)^{2}%
\bigg\}\nonumber\\
=\max\bigg\{ &  \big(\left(  (1-\eta)q_{3}+c_{3}\right)  |z|+\eta
\big)^{2},\nonumber\\
&  \big(-\left(  (1-\eta)q_{3}+c_{3}\right)  |z|+\eta\big)^{2}%
\bigg\}\nonumber\\
= &  \big(|(1-\eta)q_{3}+c_{3}||z|+\eta\big)^{2}.
\end{align}
This is clearly minimized when $c_{3}=-(1-\eta)q_{3}$. For the $x$
and $y$ terms, they are clearly minimized for
\[
c_{1}=\sqrt{1-\eta}q_{1},~c_{2}=-\sqrt{1-\eta}q_{2}.
\]
Remembering that $\mathcal{T}$ is invariant under $\sigma_{x}$,
and thus if
$(q_{1},q_{2},q_{3})$ belongs to the tetrahedron so too does $(q_{1}%
,-q_{2},-q_{3})$, therefore we can again conclude that the channel
corresponding to $(c_{1},c_{2},c_{3})$ is Pauli, and our second part of the
proposition is proved. $\square$

\begin{lemma}\label{diamondovertrace}
$||\mathcal{E}_1-\mathcal{E}_2||_\diamond\geq||\mathcal{E}_1-\mathcal{E}_2||_1$.
\end{lemma}

\noindent\textbf{Proof.}
\begin{align}
||\mathcal{E}_1-\mathcal{E}_2||_\diamond&=\sup_{\rho\in \kappa\otimes\mathcal{H}}||\mathrm{I_\kappa}\otimes\mathcal{E}_1\left(\rho\right)-\mathrm{I_\kappa}\otimes
\mathcal{E}_2\left(\rho\right)||_1\\
&\geq \sup_{\rho\in \text{sep}\left(\kappa\otimes\mathcal{H}\right)}||\mathrm{I_\kappa}\otimes\mathcal{E}_1\left(\rho\right)-\mathrm{I_\kappa}\otimes
\mathcal{E}_2\left(\rho\right)||_1\\
&=\sup_{\rho_1\otimes\rho_2}||\mathrm{I_\kappa}\otimes\mathcal{E}_1\left(\rho_1\otimes\rho_2\right)-\mathrm{I_\kappa}\otimes
\mathcal{E}_2\left(\rho_1\otimes\rho_2\right)||_1\\
&=\sup_{\rho_1\otimes\rho_2}||\rho_1\otimes\mathcal{E}_1\left(\rho_2\right)-\rho_1\otimes
\mathcal{E}_2\left(\rho_2\right)||_1\\
&=\sup_{\rho_2}||\mathcal{E}_1\left(\rho_2\right)-
\mathcal{E}_2\left(\rho_2\right)||_1\\
&=||\mathcal{E}_1-\mathcal{E}_2||_1.
\end{align}
Note we have used the property of subadditivity over tensor
product of the trace norm. $\square$

\bigskip

\noindent\textbf{Full proof of Proposition \ref{posdiamond}.}\\
 We shall begin with the case where $u=0$. First off, we know
that
\begin{equation}
\min_{\mathcal{E}_2\in\text{Pauli}}||\mathcal{E}_{\text{sim}}-\mathcal{E}_2||_\diamond\leq ||\mathcal{E}_{\text{sim}}-\mathcal{E}_{\text{cl}}||_\diamond.
\end{equation}
In order to find the diamond norm between $\mathcal{E}_\text{sim}$ and $\mathcal{E}_{\text{cl}}$, we look at
\begin{equation}
||\mathbb{I}_2\otimes\mathcal{E}_\text{sim}\left(\rho\right)-
\mathbb{I}_2\otimes\mathcal{E}_\text{cl}\left(\rho\right)||_1
\end{equation}
for an arbitrary 2 qubit state $\rho$. We find the matrix
$M_D=\left(\mathbb{I}_2\otimes\mathcal{E}_\text{sim}\left(\rho\right)-
\mathbb{I}_2\otimes\mathcal{E}_\text{cl}\left(\rho\right)\right)$
to be

\begin{equation}
M_D=\frac{1}{4}\left(\begin{array}{cccc}
(1+a_3)\eta & 0 & (a_1-\im a_2)\eta & 0\\
0 & 1(1+a_3)\eta & 0 & -(a_1- \im a_2)\eta\\
(a_1+\im a_2)\eta & 0 & (1-a_3)\eta & 0\\
0 & -(a_1+\im a_2)\eta & 0 & (-1+a_3)\eta
\end{array}\right)
\end{equation}
which has eigenvalues
\begin{align}
\frac{1}{4}\Big(-1 - &\sqrt{a_1^2+a_2^2+a_3^2}\Big)\eta &
\frac{1}{4}\Big(1 - &\sqrt{a_1^2+a_2^2+a_3^2}\Big)\eta\\
\frac{1}{4}\Big(-1 + &\sqrt{a_1^2+a_2^2+a_3^2}\Big)\eta &
\frac{1}{4}\Big(1 + &\sqrt{a_1^2+a_2^2+a_3^2}\Big)\eta.\nonumber
\intertext{Remembering that $a_1^2+a_2^2+a_3^2\leq 1$, this means the singular values are:}
\frac{1}{4}\Big(1 + &\sqrt{a_1^2+a_2^2+a_3^2}\Big)\eta  &
\frac{1}{4}\Big(1 - &\sqrt{a_1^2+a_2^2+a_3^2}\Big)\eta\\
\frac{1}{4}\Big(1 - &\sqrt{a_1^2+a_2^2+a_3^2}\Big)\eta &
\frac{1}{4}\Big(1 + &\sqrt{a_1^2+a_2^2+a_3^2}\Big)\eta,\nonumber
\end{align}
and thus their sum is $\eta$. This gives $||\mathcal{E}_{\text{sim}}-\mathcal{E}_{\text{cl}}||_\diamond=\eta=||\mathcal{E}_{\text{sim}}-\mathcal{E}_{\text{cl}}||_1$.\\
Using this, suppose there exists a channel $\mathcal{E}'$ with a
strictly smaller diamond norm than our closest channel. Then we
have the chain of inequalities
\begin{equation}
||\mathcal{E}_\text{sim}-\mathcal{E}'||_1\leq||\mathcal{E}_\text{sim}-\mathcal{E}'||_\diamond<||\mathcal{E}_\text{sim}-\mathcal{E}_\text{cl}||_\diamond=\eta=||\mathcal{E}_\text{sim}-\mathcal{E}_\text{cl}||_1,
\end{equation}
leading to a contradiction, since we know the closest channel under trace norm to be $\mathcal{E}_\text{cl}$. Thus we are forced to conclude that the diamond norm is smallest between $\mathcal{E}_\text{sim}$ and $\mathcal{E}_\text{cl}$, with distance $\eta$.\\

In the case where $u=1$,
we are writing $\mathcal{E}_\text{sim}$
as the unitary $\sigma_x$ applied after a similar simulable
channel,
$\mathcal{E}_\text{pos}=\mathcal{E}_{\eta}\circ\mathcal{E}_{P}$.
This has closest Pauli channel
$\mathcal{E}_\text{poscl}=\left(\sqrt{1-\eta}q_1,\sqrt{1-\eta}q_2,(1-\eta)q_3\right)$.
Since the trace norm is invariant under unitaries, and the 2 qubit
channel $\mathbb{I}_2\otimes\sigma_x$ is unitary, we can conclude
that
\begin{align*}
&\phantom{=}||\mathcal{E}_\text{pos}-\mathcal{E}_\text{poscl}||_\diamond \\
 &=\sup_{\rho}||\mathbb{I}_2\otimes\mathcal{E}_\text{pos}\left(\rho\right)-
\mathbb{I}_2\otimes\mathcal{E}_\text{poscl}\left(\rho\right)||_1\\
&=\sup_{\rho}||\left(\mathbb{I}_2\otimes\sigma_x\right)\left(\mathbb{I}_2\otimes\mathcal{E}_\text{pos}\left(\rho\right)-
\mathbb{I}_2\otimes\mathcal{E}_\text{poscl}\left(\rho\right)\right)||_1\\
&=\sup_{\rho}||\mathbb{I}_2\otimes\mathcal{E}_\text{sim}\left(\rho\right)-
\mathbb{I}_2\otimes\mathcal{E}_\text{cl}\left(\rho\right)||_1\\
&=||\mathcal{E}_\text{sim}-\mathcal{E}_\text{cl}||_\diamond,
\end{align*}
where we have spotted that the channel $\mathcal{E}_{\text{cl}}=\sigma_x\otimes\mathcal{E}_{\text{poscl}}$. We can then conclude that $||\mathcal{E}_\text{sim}-\mathcal{E}_\text{cl}||_\diamond=\eta$, and therefore by using the same chain of inequalities (\ref{chain}), we force this to be the minimum distance possible. $\square$


\twocolumngrid


\begin{thebibliography}{99}
\bibitem{NoteSIM} Note that, in a quantum communication scenario,
the most general type of channel simulation must be based on
LOCCs, because Alice and Bob are remote parties and may only apply
operations on their local systems. Furthermore, the simplification
of the capacity upper bound, based the relative entropy of
entanglement, exploits the requirement for trace-preserving LOCCs.
However, in other scenarios (e.g., quantum computing or quantum
metrology), we do not require remote parties and we may think of
Alice and Bob to be the same entity. In this context, one can
allow for types of channel simulation which are based on joint
quantum operations, i.e., non-local between Alice and
Bob~\cite{qsim0,qsim1,qsim2,qsim3}. It is also clear that in
general one may simulate a quantum channel by performing a
dilation into an environment which is prepared into a pure or a
mixed state~\cite{Arvind}.
\end{thebibliography}
\end{document}